\documentclass[lettersize,journal]{IEEEtran}
\usepackage{amsmath,amsfonts}
\usepackage{algorithmic}
\usepackage{algorithm}
\usepackage{array}
\usepackage[caption=false,font=normalsize,labelfont=sf,textfont=sf]{subfig}
\usepackage{textcomp}
\usepackage{stfloats}
\usepackage{url}
\usepackage{verbatim}
\usepackage{graphicx}
\usepackage{cite}
\usepackage{multirow}
\usepackage{makecell}
\usepackage{graphicx}
\usepackage{tabularx}
\usepackage{booktabs}

\begin{document}
\title{\fontsize{24pt}{28pt}\selectfont
	Channel Measurements and Modeling\\ 
	based on Composite Environmental Factor for\\ Urban Street-Canyon Intersections}

%% 作者
\author{
	Xinwen Chen,~\IEEEmembership{Student Member,~IEEE},
	~Ruisi He,~\IEEEmembership{Senior Member,~IEEE},
	~Mi Yang,~\IEEEmembership{Member,~IEEE},
	~Zhengyu Zhang,~\IEEEmembership{Student Member,~IEEE},
	~Junzhe Song,~\IEEEmembership{Student Member,~IEEE},
	~Jiahui Han,
	~Haoxiang Zhang
% <-this % stops a space

\thanks{X.~Chen, R.~He, M.~Yang, Z.~Zhang and J.~Song are with the School of Electronics and Information Engineering, Beijing Jiaotong University, Beijing 100044, China (email: 25110109@bjtu.edu.cn; ruisi.he@bjtu.edu.cn; myang@bjtu.edu.cn; 21111040@bjtu.edu.cn; 25115063@bjtu.edu.cn)
	
J.~Han and H.~Zhang are with China Academy of Industrial Internet, Ministry of Industry and Information Technology, Beijing 100015, China (email: hjh1760708@126.com; zhx61778294@126.com)
}%

}



\maketitle

%% 摘要
\begin{abstract}
	In urban environments, vehicle-to-everything (V2X) communications require accurate wireless channel characterization. This requirement is particularly critical at street-canyon intersections, where building blockage and rich multipath propagation can severely degrade link reliability. Due to its unique environmental layout, the channel characteristics in urban canyon are influenced by building distribution. However, this feature has not been well captured in existing channel models. In this paper, we propose an environment-related statistical channel model based on 5.8~GHz channel measurements. We construct a composite environmental factor to characterize environmental differences in intersections. Then, the factor is incorporated into 3GPP path-loss model and further linked to small-scale channel parameters. Finally, accuracy of the proposed model is validated using second-order channel statistics. The results show that the proposed model can effectively characterize propagation properties of urban street-canyon intersection channels with different building conditions. The proposed model provides a physically interpretable and statistically effective framework for channel simulation and performance evaluation in urban vehicular scenarios.
\end{abstract}

%% 关键词
\begin{IEEEkeywords}
Vehicular communications, urban intersections, Third Generation Partnership Project (3GPP), multipath propagation, channel model.
\end{IEEEkeywords}

\section{Introduction}
\IEEEPARstart{R}{oad} traffic safety and efficiency remain major global challenges, with road crashes and congestion causing substantial human and economic losses each year~\cite{14}. Therefore, vehicular communication technologies, commonly referred to as V2X, have been widely recognized as key enablers for improving the reliability and responsiveness of future intelligent transportation systems by establishing low-latency and highly reliable links among vehicles, roadside infrastructure, vulnerable road users, and cloud nodes~\cite{18}. In recent years, extensive efforts have been devoted to measurement-based vehicular wireless channel characterization and modeling~\cite{He2020mmwave}, covering a wide range of road environments, carrier frequencies, antenna configurations, and modeling paradigms~\cite{85,86,87}.
 Among the various road environments, urban intersections constitute safety-critical hotspots where conflicting traffic flows and building-induced non-line-of-sight (NLOS) conditions frequently coexist, making both visual perception and wireless connectivity particularly vulnerable. In such scenarios, many representative V2X applications---including collision warning, intersection movement assist, and cooperative lane-changing---strongly rely on robust links around the intersection area, which in turn require accurate and scenario-specific wireless channel models for street-canyon intersections~\cite{86}.

Among the many roadway scenarios, urban intersections in street-canyon environments are not only hotspots for traffic accidents due to the combined impact of building blockage and traffic conflicts~\cite{He2024book}, but also critical locations where V2X safety services most urgently require reliable coverage and connectivity~\cite{51}. When vehicles approach and traverse an intersection, they do so along multiple arms of the crossroads, LOS paths are frequently occluded by surrounding buildings and large vehicles, and the local propagation conditions can change from LOS to NLOS~\cite{Zheng2020nlos} and back within a distance of only a few tens of meters. As a result, the channel characteristics at intersections may differ substantially from those observed on highways~\cite{He2013Cutting},\cite{12} or along conventional urban arterials~\cite{Qian6G}, and can be highly sensitive to the specific building morphology and traffic configuration. Measurement-based channel characterization and modeling at intersections are therefore essential for accurate link budgeting, robust beam management, and reliable interference assessment in the vicinity of these safety-critical locations. This, in turn, motivates a careful evaluation of whether widely used standardized models can adequately capture intersection-specific propagation. It also highlights the need for refined formulations that explicitly account for the underlying street-canyon geometry.

For system-level simulations of 5G/6G and V2X networks, standardized channel models such as 3GPP TR~38.901~\cite{TR38901}, WINNER~II~\cite{WINNER2}, and the COST-231 Walfisch--Ikegami model~\cite{COST231} provide large-scale and small-scale statistical parameters for typical cellular scenarios, and have been widely used in macrocell, small-cell hotspot, and generic urban street environments. However, these models are largely built on an ``average urban morphology'' assumption; local variations in building layout, height distribution, and street orientation around a specific intersection are only implicitly accounted for, which makes it difficult to capture the detailed propagation behavior in dense street canyons with strong corner blockages. Recent surveys have also pointed out that, in highly heterogeneous V2V environments, a single scenario label is often insufficient to represent the impact of key geometrical factors on path loss and directional statistics~\cite{Zemen2025SSCR}.  

Dedicated studies on channel measurement and modeling for road--intersection scenarios have been reported in the literature. Early measurement campaigns were mainly conducted at 790~MHz~\cite{Sai} and 2.5/5.0~GHz~\cite{Mur},\cite{Ito}, and mostly focused on suburban~\cite{Liu} or campus-type intersections. More recent campaigns have largely moved to the 5.9~GHz band and adjacent licensed bands for vehicular communications~\cite{Gus}. The 5.8~GHz band considered in this work is also one of the widely used bands for vehicular networks. In terms of scenario types, most reported measurement sites are suburban or rural crossroads, while only a limited number of studies consider urban or dense-urban intersections surrounded by buildings. However, dense urban areas are where intersections are most common. Some studies further refine the geometric classification of intersections. For example, \cite{Rad} distinguishes between ``open'' four-way intersections with no nearby buildings and ``single-building'' intersections where one corner is occupied by a tall structure. Reference~\cite{Eva} identifies three typical conditions, namely, NLOS intersections without large reflecting surfaces, NLOS intersections with dominant reflectors, and LOS intersections with relatively unobstructed streets. Despite these efforts, most existing intersection-related studies still describe the surrounding buildings only in qualitative terms, such as ``open'' and ``closed'' intersections, and do not provide a systematic treatment of quantitative morphological parameters. In particular, for dense-urban intersections, the joint effects of building density and height variation on channel statistics have not yet been fully clarified.

In terms of channel characteristics, most existing studies still remain at the stage of channel characterization rather than channel modeling. These works report path-loss exponents, shadow-fading statistics, RMS delay and Doppler spreads, decorrelation distances, equivalent stationarity regions, and azimuth and elevation angular spreads for different intersection types~\cite{pat},\cite{Del}. The results consistently show that intersection channels exhibit clearer LOS/NLOS segmentation, larger delay and angular spreads, and shorter stationarity regions than highways~\cite{17} or straight urban roads. Based on these measurements, several cluster-based~\cite{16},\cite{19} and ray-tracing-based channel models~\cite{Bai2025RIS} have been proposed. In these models, dominant specular components and diffuse multipath components are jointly considered to reproduce the measured time--frequency evolution. Some studies have further employed array measurements to obtain AoA/EoA distributions and have reported angle spreads, time--angle evolution of dominant paths, and spatial non-stationarity under different intersection layouts and blockage conditions~\cite{5}. However, when angular characteristics are considered, the analysis often remains limited to AoA/EoA spread statistics~\cite{55}. At the modeling level, some works adopt stochastic models driven by statistical parameters, whereas others construct deterministic ray-tracing models based on detailed scene geometry~\cite{9},\cite{61}. In addition, some studies exploit vision information or digital twin~\cite{Zhang2023DT} for channel modeling, while others investigate AI-based channel prediction~\cite{He2026AI},\cite{He2025INTERACT}. Overall, these efforts cover the main methodological families currently used for channel modeling in intersection scenarios.

Despite all the efforts, the intersections are still one of the most unexplored scenarios for V2X channel modeling. Compared with the above literature, this paper proposes an environment-related channel modeling framework for urban street-canyon intersections and analyze the channel characteristics. At the large-scale level, the framework builds upon the 3GPP UMi LOS/NLOS path-loss formulas by compressing building-related statistics into a single environmental factor, which is embedded into the LOS/NLOS coefficients to yield an environment-related path-loss model. For the small-scale level, the introduced environmental factor is further linked to the small-scale channel parameters, thereby enabling the construction of a complete environment-related channel model. To this end, the main contributions of this work can be summarized as follows:

\begin{itemize}
	
	\item A statistical channel model based on composite environmental factor is proposed for urban street-canyon intersections. The factor is introduced to characterize the local intersection morphology.
	
	\item The conventional 3GPP path-loss model is extended by incorporating the environmental factor, so that the large-scale attenuation can be adaptively characterized under different intersection environments. In addition, the same environmental factor is linked to the small-scale channel parameters, which enables a unified environment-dependent channel modeling framework.
	
	\item Extensive channel measurements are conducted in urban street-canyon scenarios, and three intersections with distinct building morphologies are selected for analysis. The dataset covers dynamic LOS--NLOS transitions and provides a basis for model construction and parameter extraction.
	
	\item The proposed model is validated using independent measurement data. The results show that the updated model can more accurately reproduce the channel characteristics of intersection scenarios.
	
\end{itemize}

The remainder of this paper is organized as follows. Section~II describes the proposed environment-related model. Section~III describes the channel measurement campaign. Section~IV presents the data processing and the channel characteristics of intersections. Section~V introduces the evaluation of the model. Section~VI concludes the paper.

\section{Environment-Related Channel Model}

\begin{figure}[t]
	\centering
	\includegraphics[width=1\columnwidth]{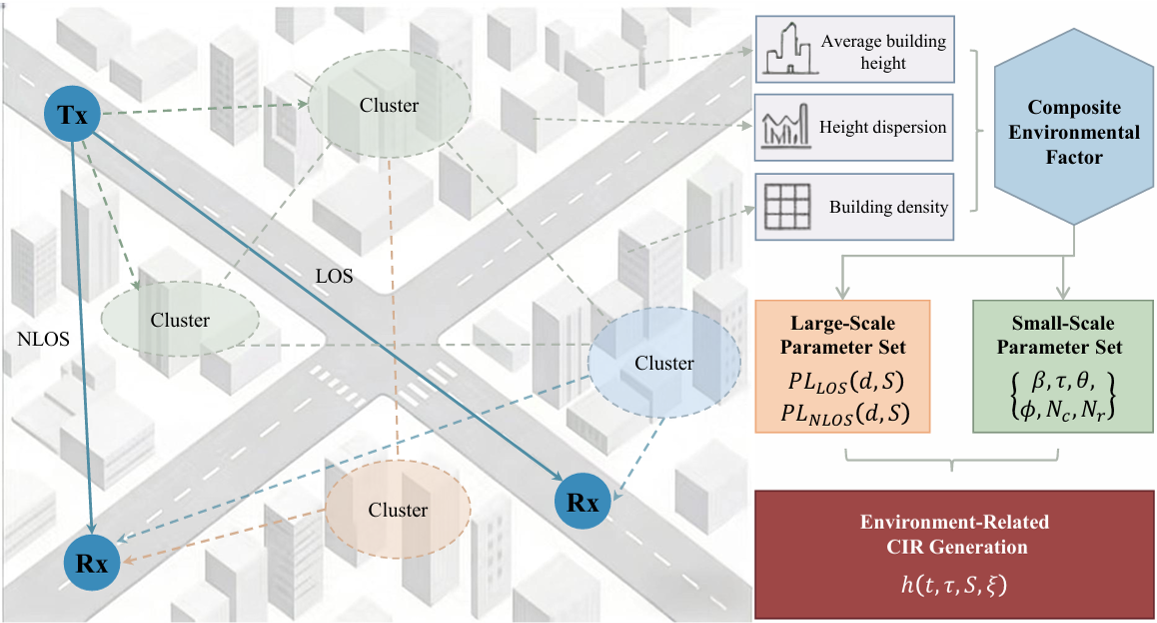}
	\caption{Schematic diagram of the environmental-related model.}
	\label{fig:model}
\end{figure}

Urban street-canyon intersections exhibit strong scenario-dependent propagation characteristics. In addition to the transmission distance and propagation condition, the height, density, and spatial distribution of roadside buildings can significantly affect signal blockage, multipath formation, and energy dispersion. As a result, both large-scale attenuation and small-scale channel statistics vary systematically with the surrounding environment. To address this issue, an environment-related statistical channel model is developed in this work, as illustrated in Fig.~\ref{fig:model}. The proposed model takes the intersection-related environmental descriptors as inputs, uses the large-scale path loss and small-scale parameters as intermediate representations, and further generates environment-related channel impulse responses, thereby establishing a unified mapping from the physical environment to the statistical channel response. The model inputs include the environmental parameters of the intersection, the transmission distance $d$, the propagation state $\xi$, and system parameters such as the carrier frequency $f_c$, where $\xi \in \{\mathrm{LOS},\mathrm{NLOS}\}$.

To quantitatively characterize the building morphology around an intersection, let $H_i$ and $s_i$ denote the height and footprint area of the $i$th building, respectively, $n$ denote the total number of buildings within the observation region, and $a$ denote the total area of the roadside observation region. Then, the average building height, the building-height dispersion, and the building density are defined as
\begin{equation}
	h_{\mathrm{height}}=\frac{\sum_{i=1}^{n}H_i s_i}{\sum_{i=1}^{n}s_i},
	\label{eq:h_height}
\end{equation}
\begin{equation}
	h_{\mathrm{std}}=\sqrt{\frac{\sum_{i=1}^{n}\left(H_i-h_{\mathrm{height}}\right)^2}{n-1}},
	\label{eq:h_std}
\end{equation}
\begin{equation}
	\rho=\frac{\sum_{i=1}^{n}s_i}{a}.
	\label{eq:rho}
\end{equation}

Here, $h_{\mathrm{height}}$ characterizes the overall building-height level of the scenario, $h_{\mathrm{std}}$ reflects the fluctuation of building heights, and $\rho$ describes the building occupancy intensity~\cite{35}. Since these three environmental descriptors jointly determine the local propagation conditions at the intersection, they are further compressed into a composite environmental factor $S$. The coefficients are fitted from the measured large-scale path-loss data, and $S$ is defined as
\begin{equation}
	S=0.5\,h_{\mathrm{height}}+0.2\,h_{\mathrm{std}}+0.8\,\rho.
	\label{eq:S}
\end{equation}

Therefore, $S$ is used as a unified environmental descriptor that links the physical intersection environment to the statistical channel parameters, providing a common input for the subsequent joint modeling of large-scale and small-scale channel characteristics.

For large-scale level, based on the 3GPP UMi LOS/NLOS structure, we embed $S$ into the model and introduce a dual-slope structure to capture the LOS--NLOS transition. The resulting environment-related formulation includes $PL_{\mathrm{LOS}}(d)$ and $PL_{\mathrm{NLOS}}(d)$. 
The modified LOS and NLOS path-loss models are expressed as

\begin{equation}
	\label{deqn_ex4}
	\begin{split}
		PL_{\mathrm{LOS}}(d) 
		= & \,(20 + k_A S)\log_{10}(d)  \\
		& + (51.4 + k_B S) + 21\log_{10}(f_c),
	\end{split}
\end{equation}

\begin{equation}
	\label{deqn_ex5}
	\begin{split}
		PL_{\mathrm{NLOS}}(d) 
		= & \,(35.3 + k_C S)\log_{10}(d) 
		+ 22.4 + 21.3\log_{10}(f_c) \\
		& - 0.3(h_{\mathrm{UT}}-1.5) 
		+ k_D S \log_{10}(d_0),
	\end{split}
\end{equation}
where $d$ denotes the Tx--Rx distance, $f_c$ is the carrier frequency, $h_{\mathrm{UT}}$ is the receiver antenna height, and $S$ is the environmental factor defined in \eqref{eq:S}. The coefficients $k_A$, $k_B$, $k_C$, and $k_D$ are empirically determined weighting factors that control the dependence of the slope and offset on $S$; their numerical values are $k_A = 0.5$, $k_B = -1.3$, $k_C = 9.1$, and $k_D = -9.2$. The parameter $d_0$ represents the breakpoint distance at which the path-loss slope changes from the LOS-dominated to the NLOS-dominated regime in the dual-slope model.

To characterize the small-scale propagation properties at urban intersections, the set of small-scale parameters is defined as
\begin{equation}
	\mathbf{X}=\{\beta,\tau,\theta,\phi,N_c,N_r\},
	\label{eq:X}
\end{equation}
where $\beta$, $\tau$, $\theta$, $\phi$, $N_c$, and $N_r$ denote the multipath power, delay, angle of arrival (AOA), elevation of arrival (EOA), number of clusters, and number of multipath components within each cluster, respectively. Since these parameters are not fixed under different intersection environments but instead exhibit evident statistical fluctuations, a conditional statistical modeling approach is adopted. Specifically, the small-scale parameters are modeled as random variables conditioned on the environmental factor $S$. For the parameter set $\mathbf{X}$, one has
\begin{equation}
	\mathbf{X}\mid S,\xi \sim F_{\mathbf{X}}\!\left(\boldsymbol{\Psi}(S,\xi)\right),
	\label{eq:X_conditional}
\end{equation}
where $F_{\mathbf{X}}(\cdot)$ denotes the conditional joint distribution of $\mathbf{X}$, and $\boldsymbol{\Psi}(S,\xi)$ is the distribution-parameter vector determined by the environmental factor $S$ and the propagation state $\xi$.

Given the environmental factor $S$, the propagation state $\xi$, and the transmission distance $d$, the large-scale path loss of the corresponding scenario is first obtained from \eqref{deqn_ex4} and \eqref{deqn_ex5}. Then, the associated small-scale parameter set is generated and the clustered multipath channel impulse response at the intersection is subsequently constructed. Then, the channel impulse response can be expressed as:
\begin{equation}
	\begin{aligned}
		h(t,\tau;S,\xi)
		&= \sum_{c=1}^{N_c(t;S,\xi)}
		\sum_{r=1}^{N_{r,c}(t;S,\xi)} e^{j\psi_{c,r}(t)} \\[6pt]
		&\quad \sqrt{10^{\beta_{c,r}(t;S,\xi)/10}\,10^{-PL_{\xi}(d(t),S)/10}} \\[6pt]
		&\quad a\bigl(\theta_{c,r}(t;S,\xi),\phi_{c,r}(t;S,\xi)\bigr)
		\delta\bigl(\tau-\tau_{c,r}(t;S,\xi)\bigr).
	\end{aligned}
\end{equation}
where the phase $\psi_{c,r}$ is randomly generated from an independent uniform distribution and is not explicitly modeled as a core statistical parameter in this work.

\section{Measurement Campaign}
This section describes the channel measurement campaign conducted in vehicular communication scenarios at urban street intersections. The collected data form the basis for the subsequent channel model development; the model structure is derived from the observed channel characteristics, and all model parameters are estimated from the measurement data.

\subsection{Measurement Equipment}
\begin{figure*}[t]
	\centering
	% 占两列的大图，宽度可以按需要微调 0.85–0.95\textwidth
	\includegraphics[width=1\textwidth]{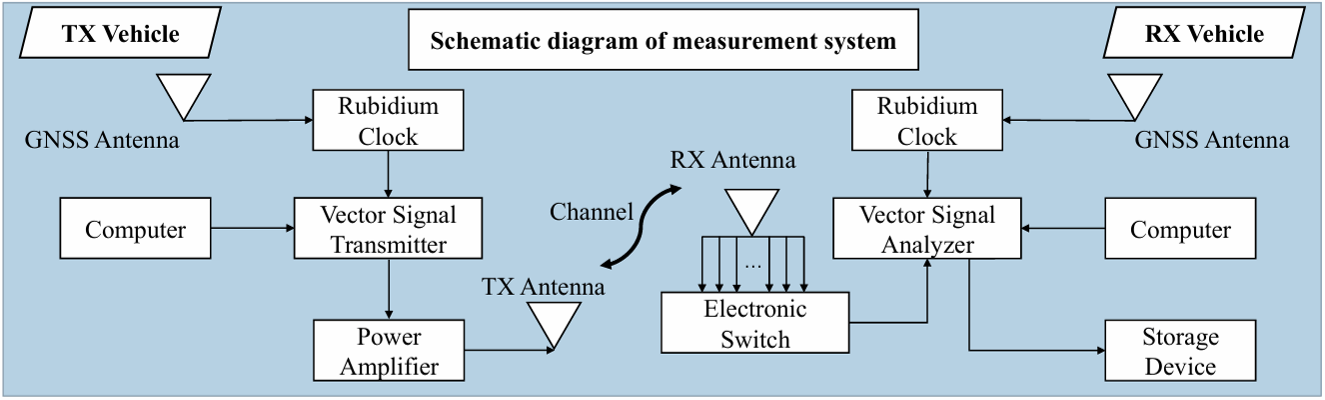}
	\captionsetup{justification=raggedright,singlelinecheck=false}
	\caption{Vehicular channel measurement system.}
	\label{fig:mea}
\end{figure*}

\begin{figure*}[t]
	\centering
	% 占两列的大图，宽度可以按需要微调 0.85–0.95\textwidth
	\includegraphics[width=1\textwidth]{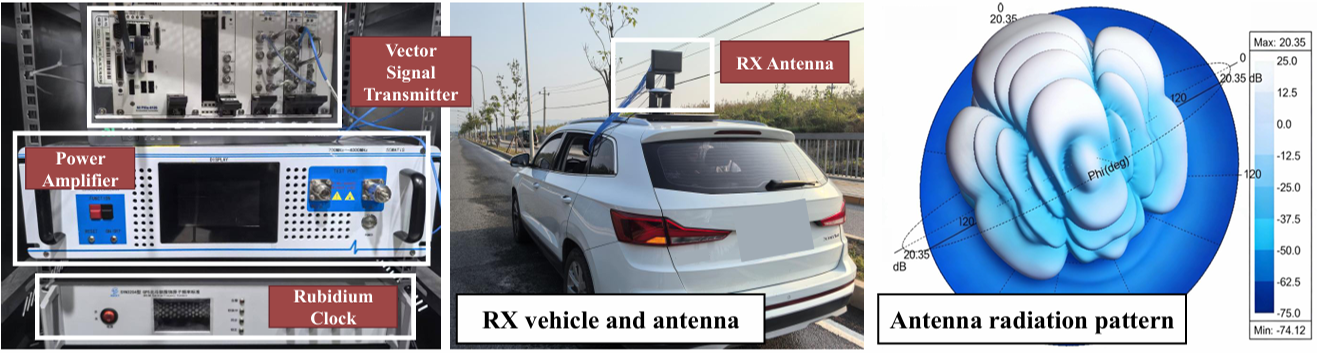}
	\caption{Main hardware components of the sounder, including the vector signal transmitter, power amplifier, rubidium clock, measurement vehicle, receive antenna array, and the three-dimensional radiation pattern of a single array element characterized in an anechoic chamber.}
	\label{fig:me}
\end{figure*}

Measurement system architecture and key equipments are shown in Fig.~\ref{fig:mea} and Fig.~\ref{fig:me}. The measurement system consists of a baseband signal source, up-conversion and power amplification modules, a receiver, measurement antennas, and synchronization and positioning units. At the transmitter (Tx), a vector signal generator produces a wideband baseband sounding signal, which is then up-converted to the carrier frequency and amplified by the RF front-end before being radiated through a roof-mounted antenna. At the receiver (Rx), a vector signal analyzer down-converts and records the received signal, from which the time- and frequency-domain channel responses are obtained via processing.

The measurement campaign is carried out at a carrier frequency of 5.8 GHz. The sounding signal is a wideband multicarrier waveform with 513 subcarriers and a total bandwidth of 30 MHz. An omnidirectional antenna is employed at the Tx, while a $4 \times 8$ antenna array is used at the Rx. Both antennas are mounted on the roofs of the measurement vehicles, with an installation height of approximately 2.5 m, which is representative of practical vehicular deployments. The (Z)-axis of the receive array is oriented opposite to the driving direction of the vehicle, providing a consistent reference for angle definition and subsequent azimuth calibration.

To obtain reliable angular information, the 3-D radiation patterns of the Tx antenna and each element of the Rx array are measured in an anechoic chamber prior to the field campaign. The right-hand side of Fig.~\ref{fig:mea} illustrates the 3D radiation pattern of a single Rx array element. Each array element is sequentially connected to the vector signal analyzer via an electronic switch, enabling multi-port signal acquisition in a time-division manner. For time synchronization, a pair of rubidium clocks disciplined by Global Navigation Satellite System (GNSS) signals are deployed at the Tx and Rx, respectively. These clocks provide a common, high-accuracy time reference for both ends of the link. In addition, the GNSS receivers output real-time longitude and latitude information, enabling precise positioning of the transmitter and receiver and supporting the subsequent distance-based and region-based statistical analysis of the channel. To remove the impact of the measurement equipment and obtain accurate channel responses, a back-to-back calibration is performed before the field measurements. In this procedure, the Tx and Rx are directly connected via coaxial cables in a laboratory environment, and the inherent amplitude and phase response of the system is recorded~\cite{yang}. During data processing, the calibration response is used to compensate the field measurements, thereby eliminating fixed distortions introduced by the instruments, cables, and RF front-ends.

The detailed configuration of the measurement system is summarized in Table~\ref{tab:meas_sys}. The effective bandwidth of 30 MHz corresponds to a delay resolution of approximately 33.3 ns, which means that only multipath components (MPCs) with propagation distance differences larger than about 10 m can be distinguished in the delay domain. For sub-6 GHz vehicular communication systems, the available bandwidth is typically in the range of 20–30 MHz, so the measurement configuration in this work is consistent with practical system settings. 

\begin{table}[!t]
	\centering
	\caption{Configurations of measurement system}
	\label{tab:meas_sys}
	\renewcommand{\arraystretch}{1.05}
	\begin{tabular}{>{\centering\arraybackslash}m{0.48\columnwidth}
			>{\centering\arraybackslash}m{0.32\columnwidth}}
		\toprule
		\textbf{Configuration} & \textbf{Description} \\
		\midrule
		Measurement scenarios & Urban intersections \\
		Center frequency, $f$ & 5.8 GHz \\
		Bandwidth, $BW$ & 30 MHz \\
		Transmit power, $P_{TX}$ & 45 dBm \\
		Delay resolution, $\Delta\tau=\tfrac{1}{BW}$ & 33.3 ns \\
		Sounding signal & Multi-carrier signal \\
		Number of frequency points & 1024 \\
		Transmitter antenna & Omnidirectional antenna \\
		Receiver antenna & $4\times8$ planar array \\
		\bottomrule
	\end{tabular}
\end{table}

\subsection{Measurement Scenarios}
The measurement campaign is conducted in Changsha, China. The considered streets are flanked by dense buildings with heights on the order of several tens of meters, forming typical urban street-canyon environments. To investigate the impact of different building configurations on vehicular channels at intersections, three representative urban street intersections are selected, denoted as Intersections 1, 2, and 3. All three are located in typical urban street environments with multi-story or high-rise buildings along both sides of the roads. However, they differ significantly in terms of building density, average height, and height variation, providing complementary cases for modeling the effect of building characteristics on path loss.

\begin{itemize}
	\item \textbf{Intersection 1.} Intersection 1 is a relatively narrow junction which we call High-Complexity Intersection (HCL). Buildings on both sides of the streets are closely aligned with the road boundaries, with small spacing between adjacent buildings, high building density, and an average height significantly greater than that in intersections 2 and 3. The road cross-section is clearly narrower. In this scenario, even when a geometric LOS path exists, the propagation is strongly influenced by wall reflections and diffractions, and the channel exhibits propagation characteristics typical of dense urban environments with strong blockage.
	
	\item \textbf{Intersection 2.} Intersection 2 is of medium size which we call Medium-Complexity Intersection (MCL). The road width is moderate, but the buildings around the junction are arranged more compactly. The number of buildings is higher than in intersection 3, and their average height is also larger, with noticeable local variations in building height. Compared with intersection 3, partial blockage is more likely to occur when vehicles approach the intersection, making this scenario representative of medium building density and height conditions.
	
	\item \textbf{Intersection 3.} Intersection 3 is a relatively large four-way intersection which we call Low-Complexity Intersection (LCL). The roads are wide, and some setback space is reserved at the four corners of the junction. The number of surrounding buildings is small, and their average height is relatively low, resulting in an overall open environment. This scenario can be regarded as a baseline case with low building density and weak blockage effects.
\end{itemize}

\begin{figure}[t]
	\centering
	% --------- 第一行：实景卫星图 (a)-(c) ---------
	\subfloat[]{%
		\includegraphics[width=0.32\linewidth]{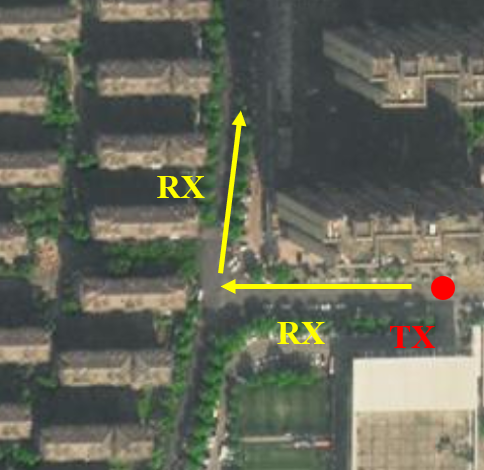}%
	}\hfil
	\subfloat[]{%
		\includegraphics[width=0.32\linewidth]{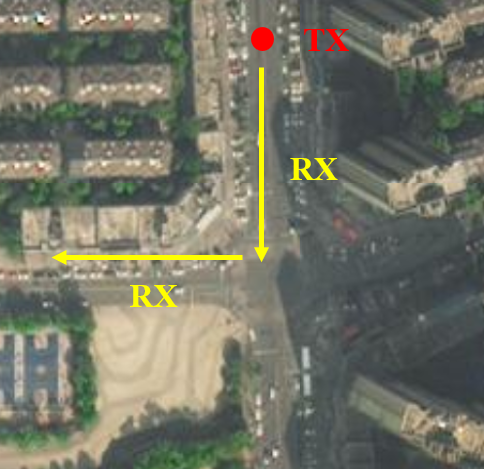}%
	}\hfil
	\subfloat[]{%
		\includegraphics[width=0.32\linewidth]{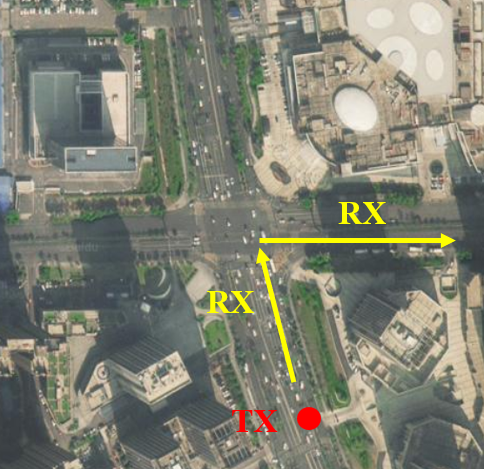}%
	}
	\caption{Measurement scenarios. (a)–(c) Aerial views of the three investigated urban intersections.}
	\label{fig:meas_scenarios}
\end{figure}

As illustrated in Fig.~\ref{fig:meas_scenarios}, the measurement campaign covers three urban four-way intersections, for which both the satellite views and the geometric positions of the TX/RX vehicles are depicted. For each intersection, one vehicle acts as the transmitter and the other as the receiver. The Tx vehicle moves along one road segment toward the intersection, while the Rx vehicle departs from an initial position ahead of the Tx, passes through the intersection, and then makes a turn, so that the link experiences a transition from LOS to NLOS. The channel state is labeled as LOS when both vehicles are on the same road segment and there is no building blocking the direct path. When the vehicles are located on different road segments and the direct path is obstructed by corner buildings, the link is classified as NLOS. By combining GNSS-based position trajectories with video recordings, the approximate spatial regions corresponding to LOS and NLOS can be determined for each scenario. During the measurements, the vehicle speed is kept at approximately 30 km/h, which reflects typical urban driving conditions while still ensuring a sufficiently high temporal sampling density of the channel. According to the magnitude of the environmental factor~$S$, the routes corresponding to Intersections~1--3 are classified as HCL, MCL, and LCL, with $S$ ranging from 40 to 50, 25 to 35, and 10 to 20, respectively. In the subsequent analysis, the midpoint of each range is adopted as the representative value of~$S$.

\section{Data Processing And Characteristic Analysis}
This section processes and analyzes the channel characteristics of urban street-canyon intersections based on the measured data described in Section~III, with the aim of providing parameter support for the environment-related statistical channel model established in Section~II. Specifically, the large-scale path-loss characteristics under different propagation conditions are first extracted and analyzed to reveal the impact of the building environment on the average attenuation behavior. Then, several key small-scale parameters are extracted and statistically analyzed, and their relationships with the environmental factor~$S$ are further investigated. Through this analysis, this section lays the foundation for the subsequent generation and validation of the environment-related statistical channel model.

\subsection{Large-Scale Characteristics}

\begin{figure*}[t]
	\centering
	\includegraphics[width=1\textwidth]{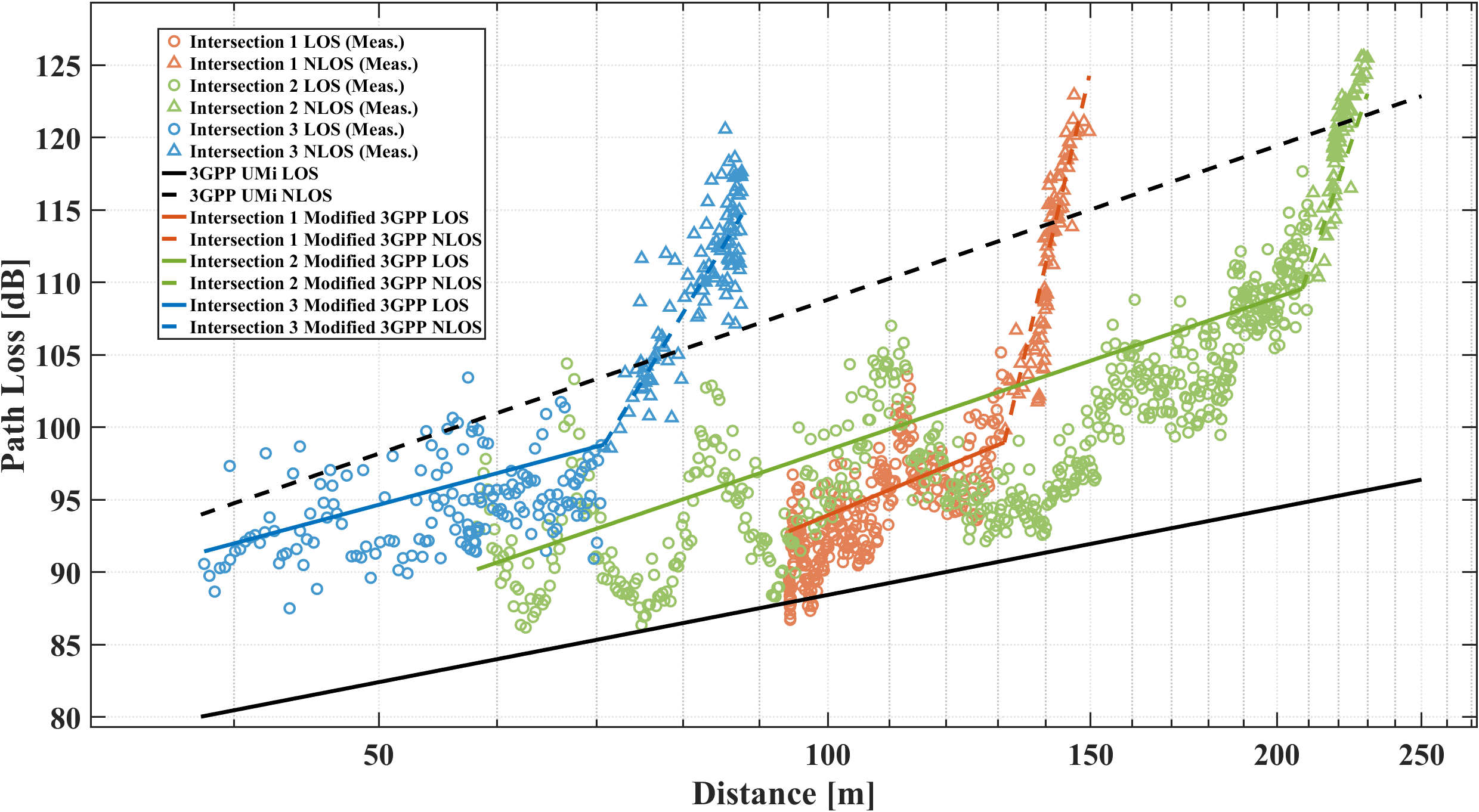}
	\caption{Measured distance-dependent path loss and linear fitting of the modified 3GPP-based path-loss model. Circles and triangles denote the measured LOS and NLOS samples, respectively. The black solid and dashed lines represent the LOS and NLOS fitting results of the 3GPP UMi model, while the remaining solid and dashed curves correspond to the LOS and NLOS fits of the proposed modified path-loss model.}
	\label{fig:pathloss_three_routes}
\end{figure*}

Path loss quantifies the attenuation between the transmitter and receiver and serves as a fundamental metric for characterizing the radio channel. In this work, it is obtained from the estimated channel transfer function as

\begin{equation}
	PL=-10\log_{10}\left(\frac{1}{N_f}\sum_{l=1}^{N_f}\lvert H(f_l)\rvert^2\right),
\end{equation}
where $N_f$ denotes the number of frequency samples used to compute the path loss and $H(f_l)$ is the frequency-domain channel response.

Fig.~\ref{fig:pathloss_three_routes} presents the measured distance-dependent path loss together with the corresponding linear fitting results. The scattered markers represent the path-loss values obtained from individual channel snapshots, while the black curves show the 3GPP UMi reference path-loss model. As can be observed, in the urban intersection scenarios the path loss increases approximately linearly with the logarithm of the Tx--Rx distance. However, due to markedly different building-blockage conditions, the growth rates in the LOS and NLOS regions are significantly different. In addition, the 3GPP model exhibits considerable systematic deviations from the measurements at different intersections and thus fails to accurately capture the slope and intercept of the measured path-loss curves.

\begin{figure}[t]
	\centering
	\includegraphics[width=1\columnwidth]{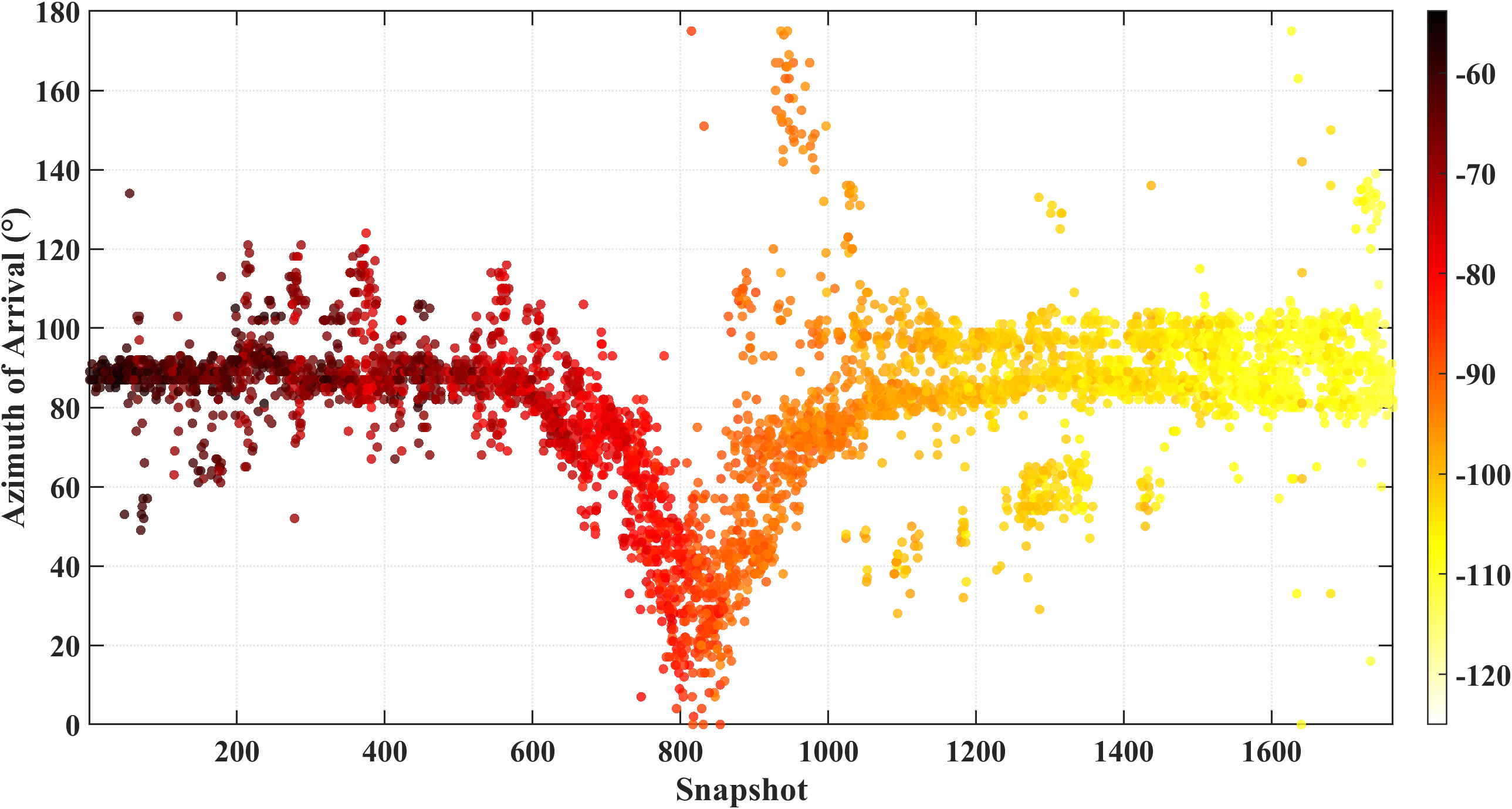}
	\captionsetup{justification=raggedright,singlelinecheck=false}
	\caption{Estimated MPC parameters.}
	\label{fig:MPC}
\end{figure}

As shown in Fig.~\ref{fig:pathloss_three_routes}, the standardized 3GPP UMi model shows clear systematic bias in street-canyon intersections. In particular, in LOS the 3GPP curve stays well below the measured samples, and the mismatch increases with distance. This indicates that the baseline LOS model cannot capture the extra loss caused by the joint effects of blockage and scattering in dense intersections. In NLOS, the 3GPP prediction may pass through the measurement cloud in some distance ranges, as also reported in~\cite{58}, but its single-slope form cannot match both the slope and the intercept across different routes. The measurements also show that the NLOS loss is strongly related to the breakpoint behavior around the LOS--NLOS transition, which is not explicitly described in the original model. By embedding the environmental factor $S$ into the LOS/NLOS coefficients and introducing a dual-slope structure, the proposed model produces path-loss slopes and offsets that are consistent with the local building morphology. On the three routes, the proposed model reduces the RMSE by about 8~dB in LOS and 3~dB in NLOS compared with the original 3GPP model. These results indicate that $S$ provides a simple and interpretable way to represent intersection-specific environment differences while keeping a 3GPP-compatible form.

\begin{table*}[t]
	\centering
	\caption{LOS state: Environment-Related Model Parameters}
	\renewcommand{\arraystretch}{1.8}
	\setlength{\tabcolsep}{3pt}
	\small
	\begin{tabular}{|>{\centering\arraybackslash}m{2.10cm}|
			>{\centering\arraybackslash}m{1.35cm}|
			>{\centering\arraybackslash}m{1.95cm}|
			>{\centering\arraybackslash}m{5.20cm}|
			>{\raggedright\arraybackslash}m{4.85cm}|}
		\hline
		$X$ & $F_X$ & Parameter & $\psi_X(S)$ & Value \\
		\hline
		
		\multirow{2}{*}{\makecell[c]{Power $\beta$ [dB]}}
		& \multirow{2}{*}{Normal}
		& $\mu_{\beta}^{\mathrm{LOS}}$
		& $\mu_{\beta}^{\mathrm{LOS}}=a_{\beta,1}^{\mathrm{LOS}}\tilde S+a_{\beta,0}^{\mathrm{LOS}}$
		& $a_{\beta,1}^{\mathrm{LOS}}=0.74,\quad a_{\beta,0}^{\mathrm{LOS}}=-6.93$ \\
		\cline{3-5}
		& & $\sigma_{\beta}^{\mathrm{LOS}}$
		& $\sigma_{\beta}^{\mathrm{LOS}}=\alpha_{\beta}^{\mathrm{LOS}}e^{\beta_{\beta}^{\mathrm{LOS}}\tilde S}$
		& $\alpha_{\beta}^{\mathrm{LOS}}=3.76,\quad \beta_{\beta}^{\mathrm{LOS}}=-0.03$ \\
		\hline
		
		\multirow{2}{*}{\makecell[c]{Delay $\tau$ [ns]}}
		& \multirow{2}{*}{Lognormal}
		& $\mu_{\log,\tau}^{\mathrm{LOS}}$
		& $\mu_{\log,\tau}^{\mathrm{LOS}}=a_{\tau,1}^{\mathrm{LOS}}\tilde S+a_{\tau,0}^{\mathrm{LOS}}$
		& $a_{\tau,1}^{\mathrm{LOS}}=-0.03,\quad a_{\tau,0}^{\mathrm{LOS}}=9.49$ \\
		\cline{3-5}
		& & $\sigma_{\log,\tau}^{\mathrm{LOS}}$
		& $\sigma_{\log,\tau}^{\mathrm{LOS}}=c_{\tau,1}^{\mathrm{LOS}}\tilde S+c_{\tau,0}^{\mathrm{LOS}}$
		& $c_{\tau,1}^{\mathrm{LOS}}=-0.0015,\quad c_{\tau,0}^{\mathrm{LOS}}=0.0195$ \\
		\hline
		
		\multirow{2}{*}{AoA $\theta$ [$^\circ$]}
		& \multirow{2}{*}{Laplace}
		& $\mu_{\theta}^{\mathrm{LOS}}$
		& $\mu_{\theta}^{\mathrm{LOS}}=c_{\theta}^{\mathrm{LOS}}$
		& $c_{\theta}^{\mathrm{LOS}}=91$ \\
		\cline{3-5}
		& & $b_{\theta}^{\mathrm{LOS}}$
		& $b_{\theta}^{\mathrm{LOS}}=\dfrac{22.62+7.21\tilde S}{\sqrt{2}}$
		& $b_{\theta}^{\mathrm{LOS}}=\dfrac{22.62+7.21\tilde S}{\sqrt{2}}$ \\
		\hline
		
		\multirow{2}{*}{EoA $\varphi$ [$^\circ$]}
		& \multirow{2}{*}{Laplace}
		& $\mu_{\varphi}^{\mathrm{LOS}}$
		& $\mu_{\varphi}^{\mathrm{LOS}}=c_{\varphi}^{\mathrm{LOS}}$
		& $c_{\varphi}^{\mathrm{LOS}}=88$ \\
		\cline{3-5}
		& & $b_{\varphi}^{\mathrm{LOS}}$
		& $b_{\varphi}^{\mathrm{LOS}}=a_{\varphi,1}^{\mathrm{LOS}}\tilde S+a_{\varphi,0}^{\mathrm{LOS}}$
		& $a_{\varphi,1}^{\mathrm{LOS}}=1.21,\quad a_{\varphi,0}^{\mathrm{LOS}}=7.31$ \\
		\hline
		
		\multirow{2}{*}{$N_{\mathrm{cl}}$}
		& \multirow{2}{*}{Normal}
		& $\mu_{N_{\mathrm{cl}}}^{\mathrm{LOS}}$
		& $\mu_{N_{\mathrm{cl}}}^{\mathrm{LOS}}=a_{cl,1}^{\mathrm{LOS}}\tilde S+a_{cl,0}^{\mathrm{LOS}}$
		& $a_{cl,1}^{\mathrm{LOS}}=0.13,\quad a_{cl,0}^{\mathrm{LOS}}=1.69$ \\
		\cline{3-5}
		& & $\sigma_{N_{\mathrm{cl}}}^{\mathrm{LOS}}$
		& $\sigma_{N_{\mathrm{cl}}}^{\mathrm{LOS}}=\alpha_{cl}^{\mathrm{LOS}}e^{\beta_{cl}^{\mathrm{LOS}}\tilde S}$
		& $\alpha_{cl}^{\mathrm{LOS}}=0.80,\quad \beta_{cl}^{\mathrm{LOS}}=0.12$ \\
		\hline
		
		\multirow{2}{*}{$N_{\mathrm{MPC}}$}
		& \multirow{2}{*}{Normal}
		& $\mu_{N_{\mathrm{MPC}}}^{\mathrm{LOS}}$
		& $\mu_{N_{\mathrm{MPC}}}^{\mathrm{LOS}}=a_{m,1}^{\mathrm{LOS}}\tilde S+a_{m,0}^{\mathrm{LOS}}$
		& $a_{m,1}^{\mathrm{LOS}}=-0.03,\quad a_{m,0}^{\mathrm{LOS}}=14.62$ \\
		\cline{3-5}
		& & $\sigma_{N_{\mathrm{MPC}}}^{\mathrm{LOS}}$
		& $\sigma_{N_{\mathrm{MPC}}}^{\mathrm{LOS}}=\alpha_{m}^{\mathrm{LOS}}e^{\beta_{m}^{\mathrm{LOS}}\tilde S}$
		& $\alpha_{m}^{\mathrm{LOS}}=0.63,\quad \beta_{m}^{\mathrm{LOS}}=0.15$ \\
		\hline
	\end{tabular}
	\label{tab:los_statistical_parameters}
\end{table*}

\subsection{Small-Scale Characteristics}

\begin{figure}[!t]
	\centering
	
	\begin{minipage}[t]{0.47\linewidth}
		\centering
		\includegraphics[width=\linewidth]{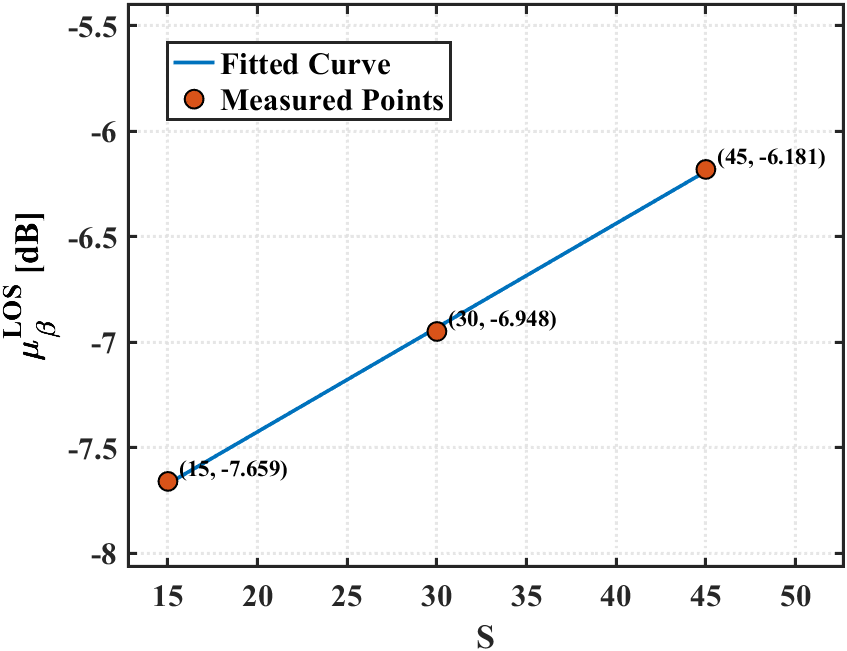}
	\end{minipage}
	\hfill
	\begin{minipage}[t]{0.47\linewidth}
		\centering
		\includegraphics[width=\linewidth]{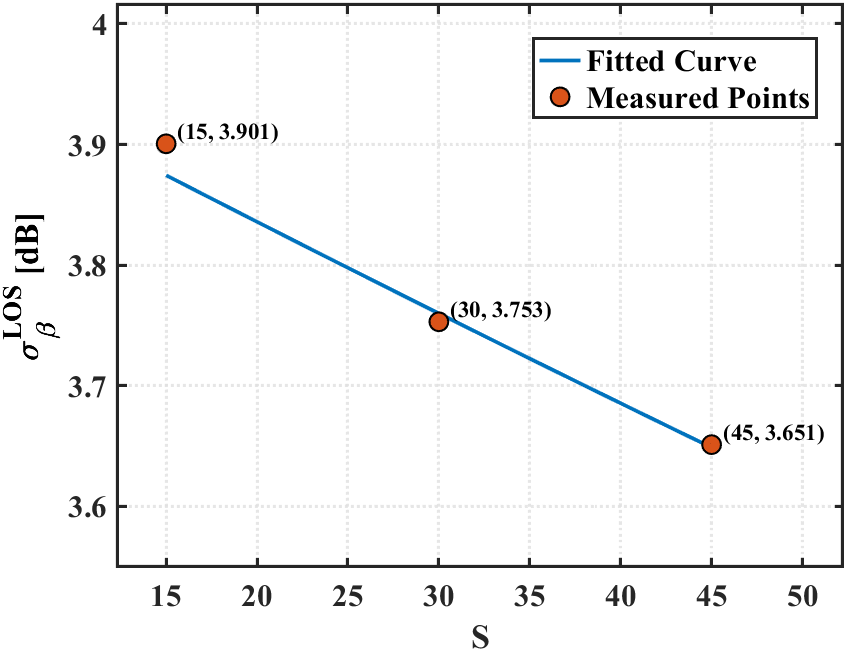}
	\end{minipage}
	
	\vspace{2mm}
	
	\begin{minipage}[t]{0.47\linewidth}
		\centering
		\includegraphics[width=\linewidth]{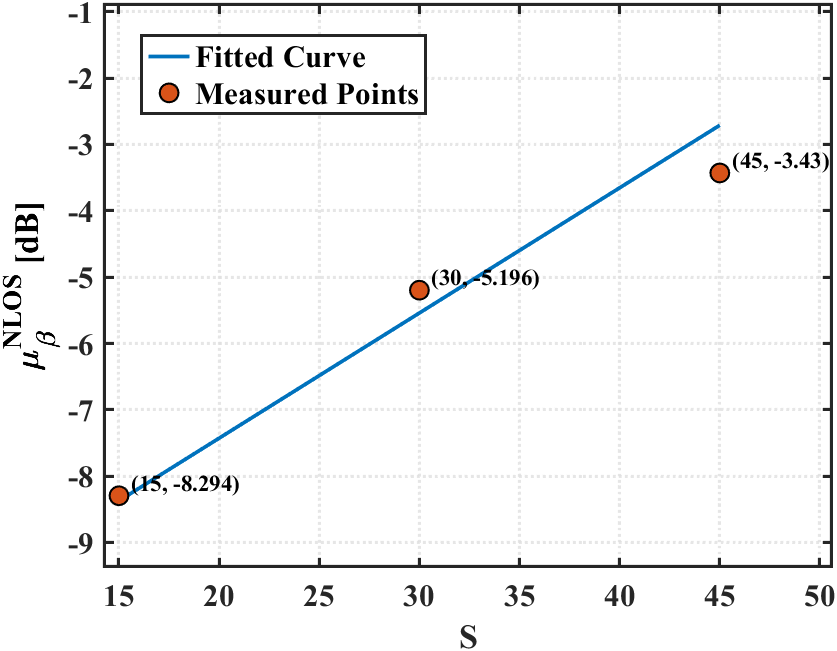}
	\end{minipage}
	\hfill
	\begin{minipage}[t]{0.47\linewidth}
		\centering
		\includegraphics[width=\linewidth]{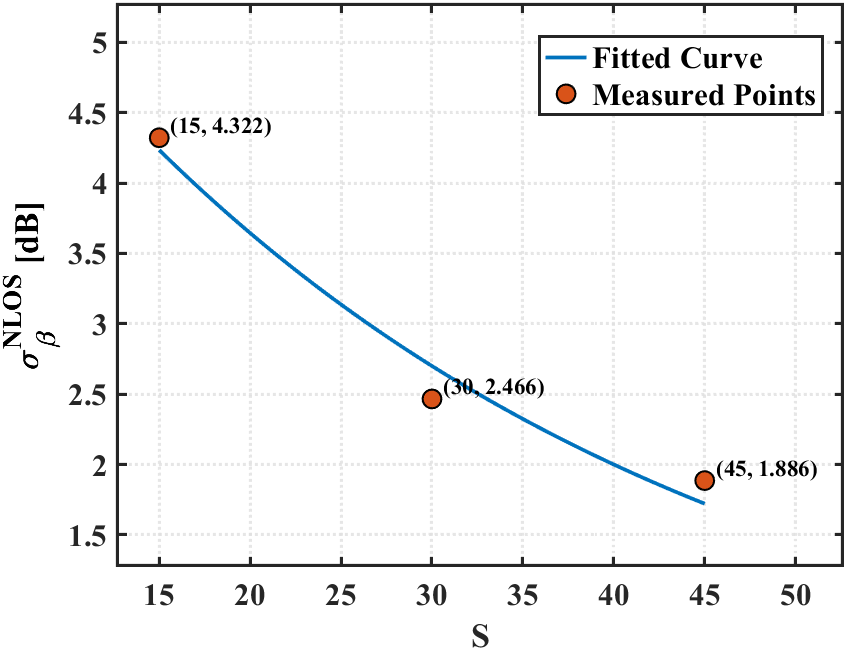}
	\end{minipage}
	
	\caption{The fitting results of the multipath power parameters.}
	\label{fig:semantic_fit}
\end{figure}

\begin{table*}[t]
	\centering
	\caption{NLOS state: Environment-Related Model Parameters}
	\renewcommand{\arraystretch}{1.8}
	\setlength{\tabcolsep}{3pt}
	\small
	\begin{tabular}{|>{\centering\arraybackslash}m{2.10cm}|
			>{\centering\arraybackslash}m{1.35cm}|
			>{\centering\arraybackslash}m{1.95cm}|
			>{\centering\arraybackslash}m{5.20cm}|
			>{\raggedright\arraybackslash}m{4.85cm}|}
		\hline
		$X$ & $F_X$ & Parameter & $\psi_X(S)$ & Value \\
		\hline
		
		\multirow{2}{*}{\makecell[c]{Power $\beta$ [dB]}}
		& \multirow{2}{*}{Normal}
		& $\mu_{\beta}^{\mathrm{NLOS}}$
		& $\mu_{\beta}^{\mathrm{NLOS}}=a_{\beta,1}^{\mathrm{NLOS}}\tilde S+a_{\beta,0}^{\mathrm{NLOS}}$
		& $a_{\beta,1}^{\mathrm{NLOS}}=2.83,\quad a_{\beta,0}^{\mathrm{NLOS}}=-5.54$ \\
		\cline{3-5}
		& & $\sigma_{\beta}^{\mathrm{NLOS}}$
		& $\sigma_{\beta}^{\mathrm{NLOS}}=\alpha_{\beta}^{\mathrm{NLOS}}e^{\beta_{\beta}^{\mathrm{NLOS}}\tilde S}$
		& $\alpha_{\beta}^{\mathrm{NLOS}}=2.70,\quad \beta_{\beta}^{\mathrm{NLOS}}=-0.45$ \\
		\hline
		
		\multirow{2}{*}{\makecell[c]{Delay $\tau$ [ns]}}
		& \multirow{2}{*}{Laplace}
		& $\mu_{\tau}^{\mathrm{NLOS}}$
		& $\mu_{\tau}^{\mathrm{NLOS}}=a_{\tau,1}^{\mathrm{NLOS}}\tilde S+a_{\tau,0}^{\mathrm{NLOS}}$
		& $a_{\tau,1}^{\mathrm{NLOS}}=-1100.00,\quad a_{\tau,0}^{\mathrm{NLOS}}=12855.50$ \\
		\cline{3-5}
		& & $b_{\tau}^{\mathrm{NLOS}}$
		& $b_{\tau}^{\mathrm{NLOS}}=\alpha_{\tau}^{\mathrm{NLOS}}e^{\beta_{\tau}^{\mathrm{NLOS}}\tilde S}$
		& $\alpha_{\tau}^{\mathrm{NLOS}}=233.80,\quad \beta_{\tau}^{\mathrm{NLOS}}=1.26$ \\
		\hline
		
		\multirow{2}{*}{AoA $\theta$ [$^\circ$]}
		& \multirow{2}{*}{Laplace}
		& $\mu_{\theta}^{\mathrm{NLOS}}$
		& $\mu_{\theta}^{\mathrm{NLOS}}=c_{\theta}^{\mathrm{NLOS}}$
		& $c_{\theta}^{\mathrm{NLOS}}=92.00$ \\
		\cline{3-5}
		& & $b_{\theta}^{\mathrm{NLOS}}$
		& $b_{\theta}^{\mathrm{NLOS}}=\alpha_{\theta}^{\mathrm{NLOS}}e^{\beta_{\theta}^{\mathrm{NLOS}}\tilde S}$
		& $\alpha_{\theta}^{\mathrm{NLOS}}=12.39,\quad \beta_{\theta}^{\mathrm{NLOS}}=0.06$ \\
		\hline
		
		\multirow{2}{*}{EoA $\varphi$ [$^\circ$]}
		& \multirow{2}{*}{Laplace}
		& $\mu_{\varphi}^{\mathrm{NLOS}}$
		& $\mu_{\varphi}^{\mathrm{NLOS}}=c_{\varphi}^{\mathrm{NLOS}}$
		& $c_{\varphi}^{\mathrm{NLOS}}=88.00$ \\
		\cline{3-5}
		& & $b_{\varphi}^{\mathrm{NLOS}}$
		& $b_{\varphi}^{\mathrm{NLOS}}=a_{\varphi,1}^{\mathrm{NLOS}}\tilde S+a_{\varphi,0}^{\mathrm{NLOS}}$
		& $a_{\varphi,1}^{\mathrm{NLOS}}=2.45,\quad a_{\varphi,0}^{\mathrm{NLOS}}=10.55$ \\
		\hline
		
		\multirow{2}{*}{$N_{\mathrm{cl}}$}
		& \multirow{2}{*}{Normal}
		& $\mu_{N_{\mathrm{cl}}}^{\mathrm{NLOS}}$
		& $\mu_{N_{\mathrm{cl}}}^{\mathrm{NLOS}}=a_{cl,1}^{\mathrm{NLOS}}\tilde S+a_{cl,0}^{\mathrm{NLOS}}$
		& $a_{cl,1}^{\mathrm{NLOS}}=0.50,\quad a_{cl,0}^{\mathrm{NLOS}}=2.70$ \\
		\cline{3-5}
		& & $\sigma_{N_{\mathrm{cl}}}^{\mathrm{NLOS}}$
		& $\sigma_{N_{\mathrm{cl}}}^{\mathrm{NLOS}}=\alpha_{cl}^{\mathrm{NLOS}}e^{\beta_{cl}^{\mathrm{NLOS}}\tilde S}$
		& $\alpha_{cl}^{\mathrm{NLOS}}=1.03,\quad \beta_{cl}^{\mathrm{NLOS}}=0.44$ \\
		\hline
		
		\multirow{2}{*}{$N_{\mathrm{MPC}}$}
		& \multirow{2}{*}{Normal}
		& $\mu_{N_{\mathrm{MPC}}}^{\mathrm{NLOS}}$
		& $\mu_{N_{\mathrm{MPC}}}^{\mathrm{NLOS}}=a_{m,1}^{\mathrm{NLOS}}\tilde S+a_{m,0}^{\mathrm{NLOS}}$
		& $a_{m,1}^{\mathrm{NLOS}}=0.06,\quad a_{m,0}^{\mathrm{NLOS}}=14.66$ \\
		\cline{3-5}
		& & $\sigma_{N_{\mathrm{MPC}}}^{\mathrm{NLOS}}$
		& $\sigma_{N_{\mathrm{MPC}}}^{\mathrm{NLOS}}=\alpha_{m}^{\mathrm{NLOS}}e^{\beta_{m}^{\mathrm{NLOS}}\tilde S}$
		& $\alpha_{m}^{\mathrm{NLOS}}=0.61,\quad \beta_{m}^{\mathrm{NLOS}}=0.01$ \\
		\hline
	\end{tabular}
	\label{tab:nlos_statistical_parameters}
\end{table*}

\begin{figure*}[!t]
	\centering
	
	\begin{minipage}[t]{0.47\textwidth}
		\centering
		{\small \textbf{LOS}}\\[2mm]
		
		\includegraphics[width=0.49\linewidth]{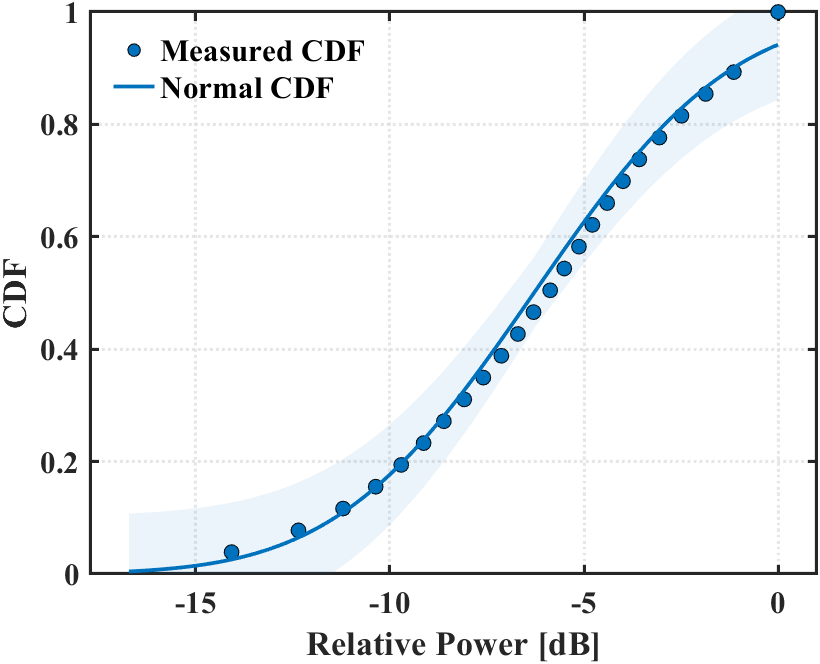}
		\includegraphics[width=0.49\linewidth]{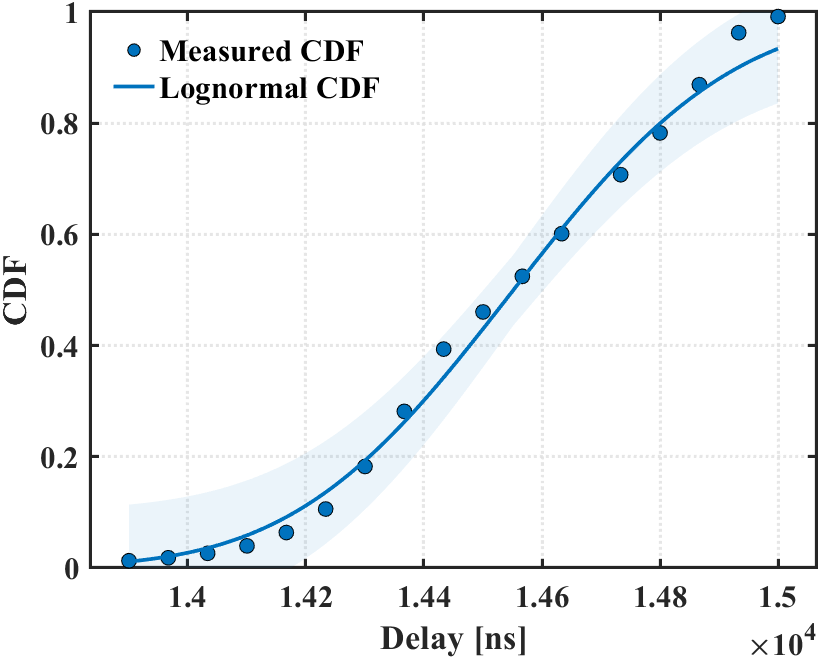}\\[1mm]
		
		\includegraphics[width=0.49\linewidth]{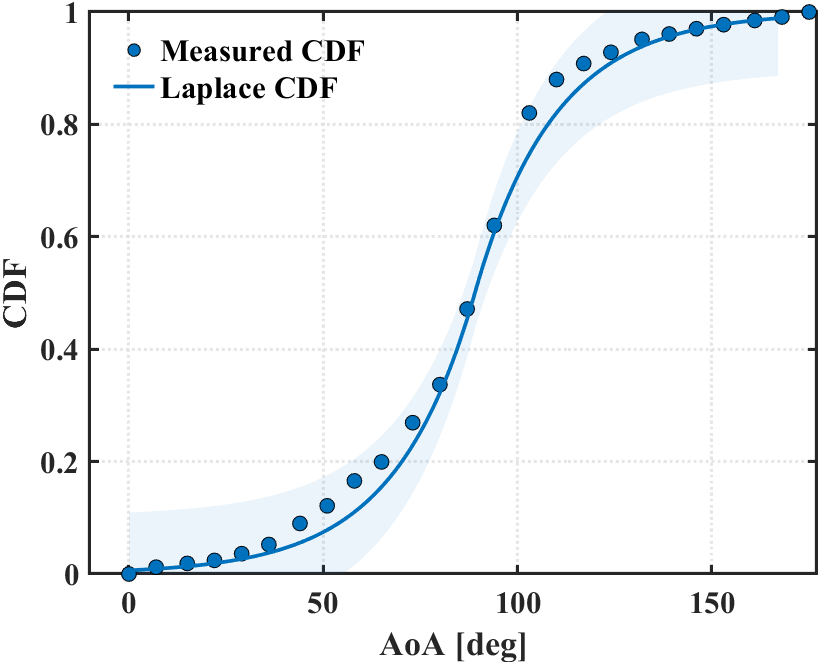}
		\includegraphics[width=0.49\linewidth]{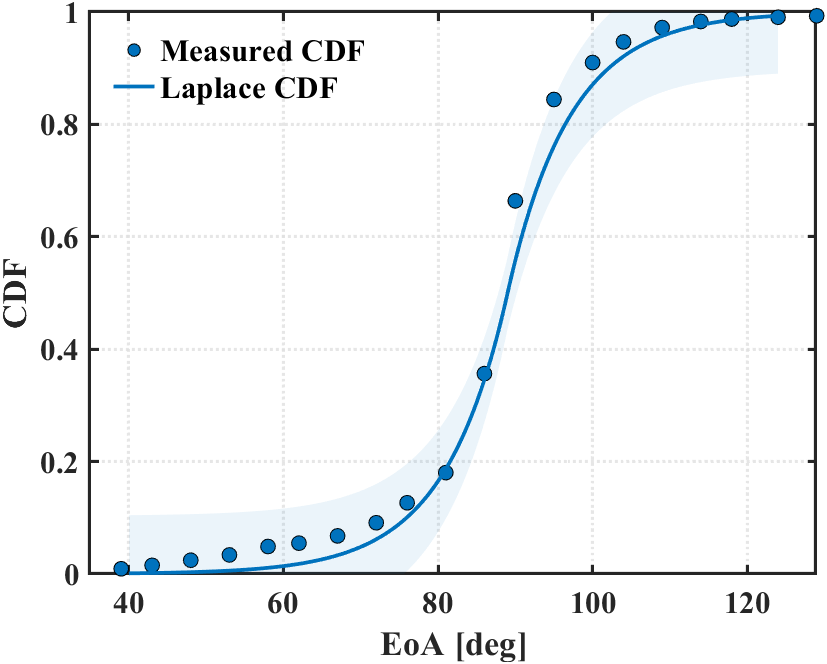}\\[1mm]
		
		\includegraphics[width=0.49\linewidth]{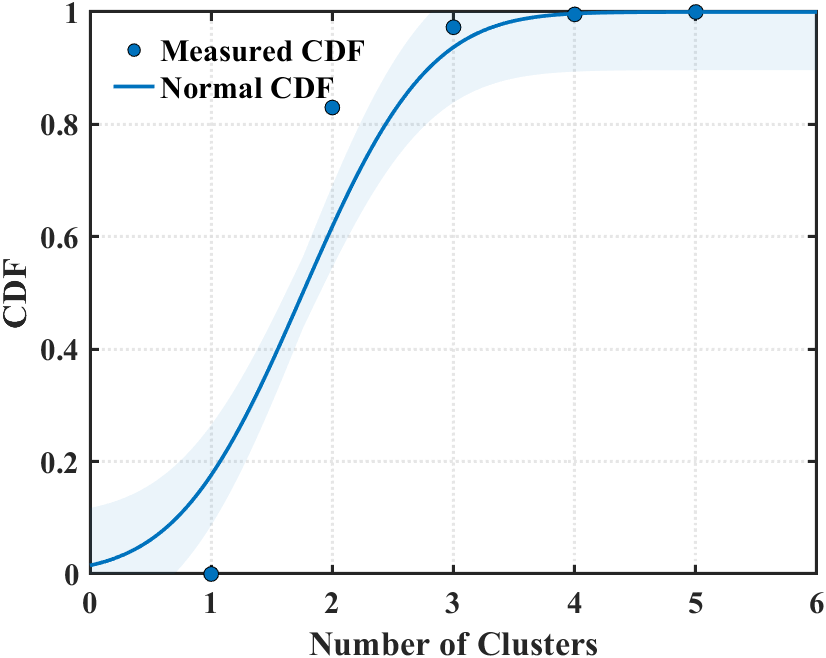}
		\includegraphics[width=0.49\linewidth]{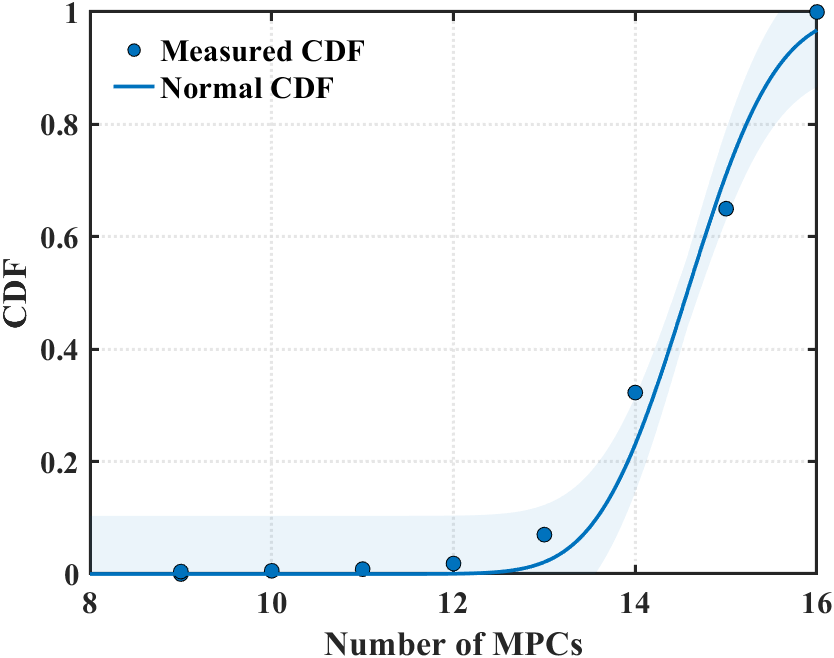}
	\end{minipage}
	\hspace{0.01\textwidth}
	\vrule width 0.5pt
	\hspace{0.01\textwidth}
	\begin{minipage}[t]{0.47\textwidth}
		\centering
		{\small \textbf{NLOS}}\\[2mm]
		
		\includegraphics[width=0.49\linewidth]{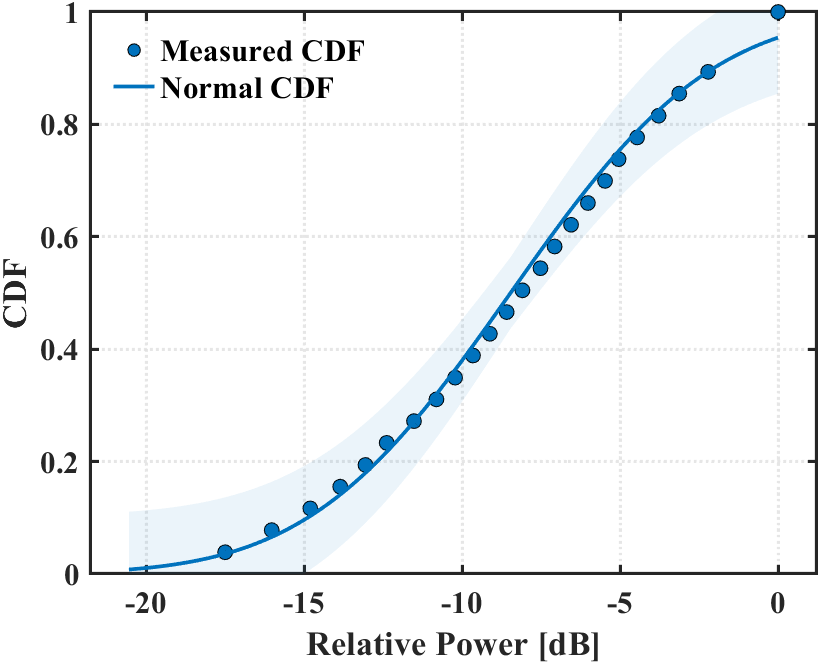}
		\includegraphics[width=0.49\linewidth]{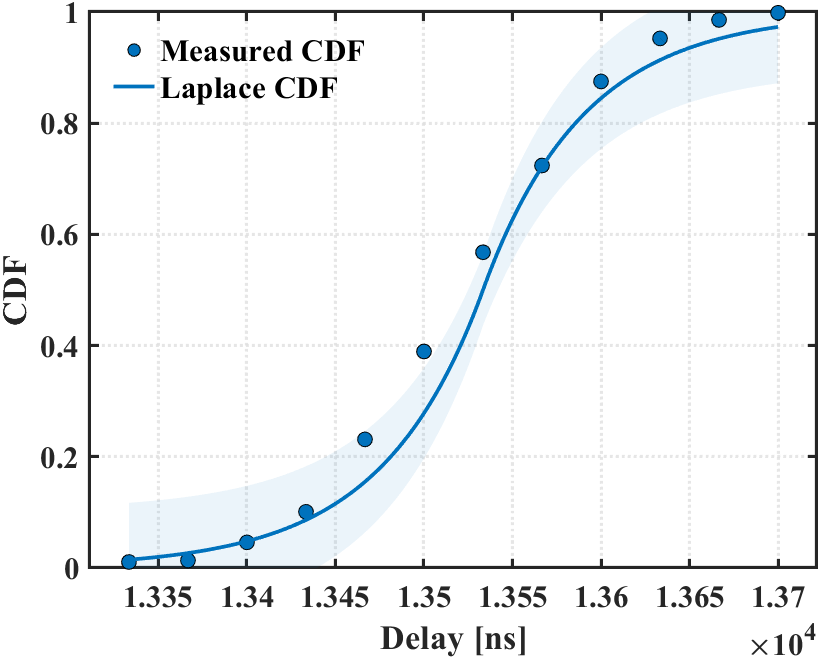}\\[1mm]
		
		\includegraphics[width=0.49\linewidth]{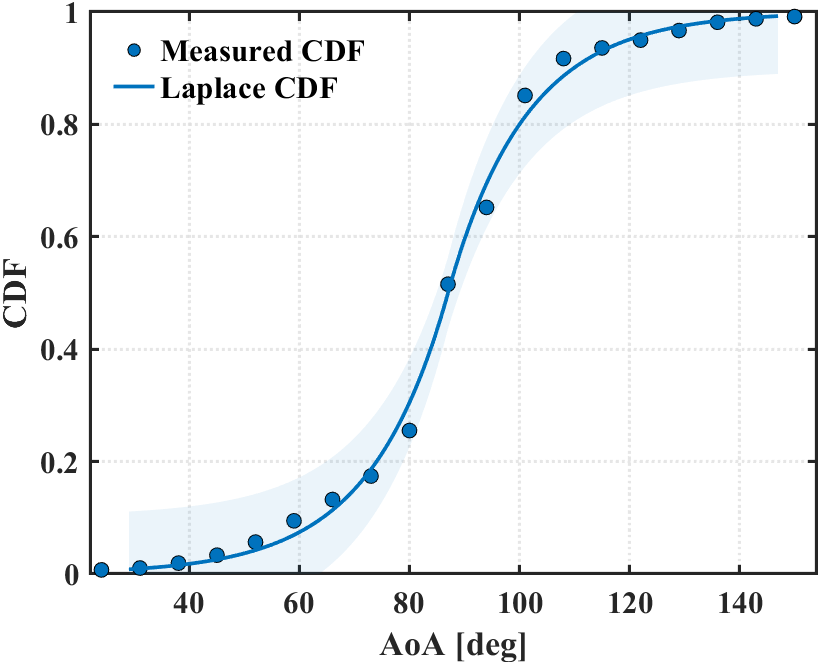}
		\includegraphics[width=0.49\linewidth]{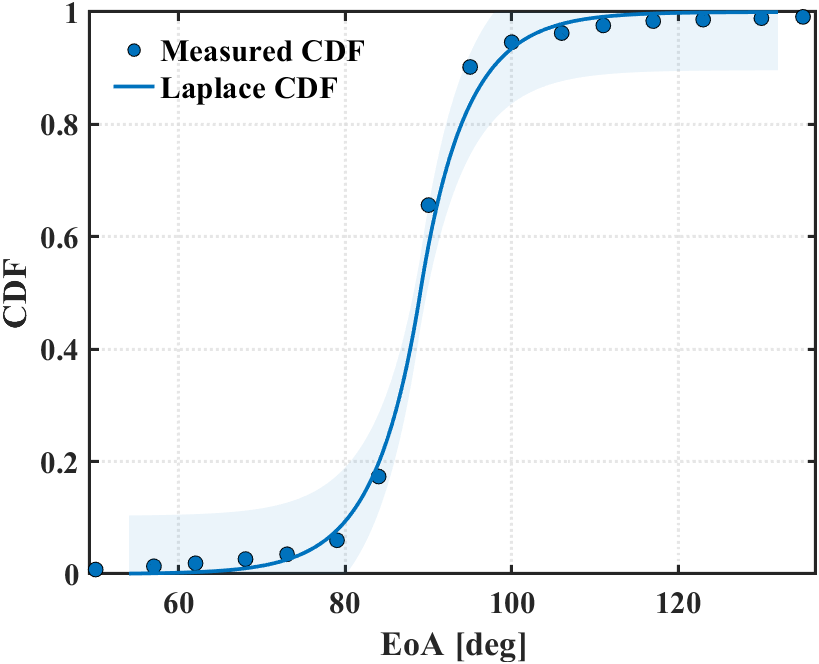}\\[1mm]
		
		\includegraphics[width=0.49\linewidth]{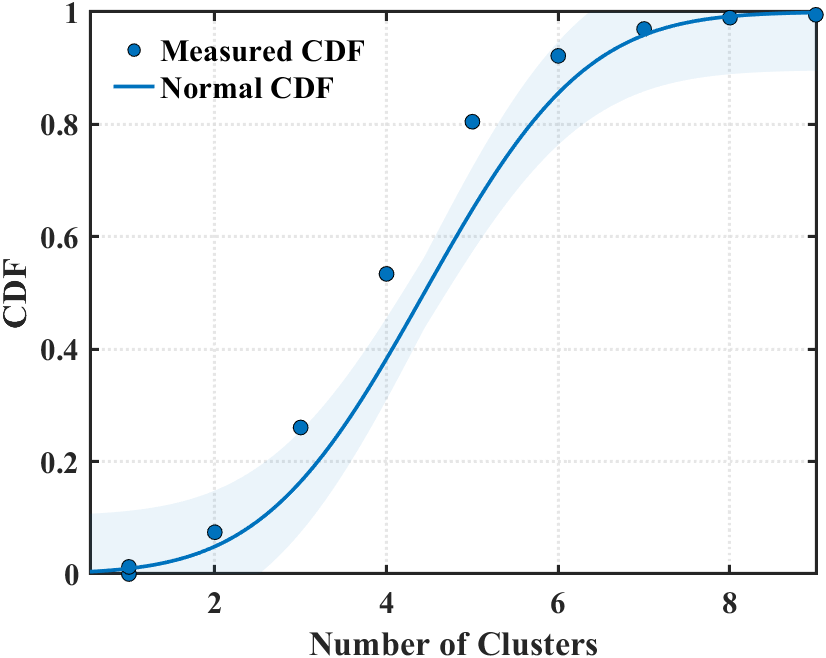}
		\includegraphics[width=0.49\linewidth]{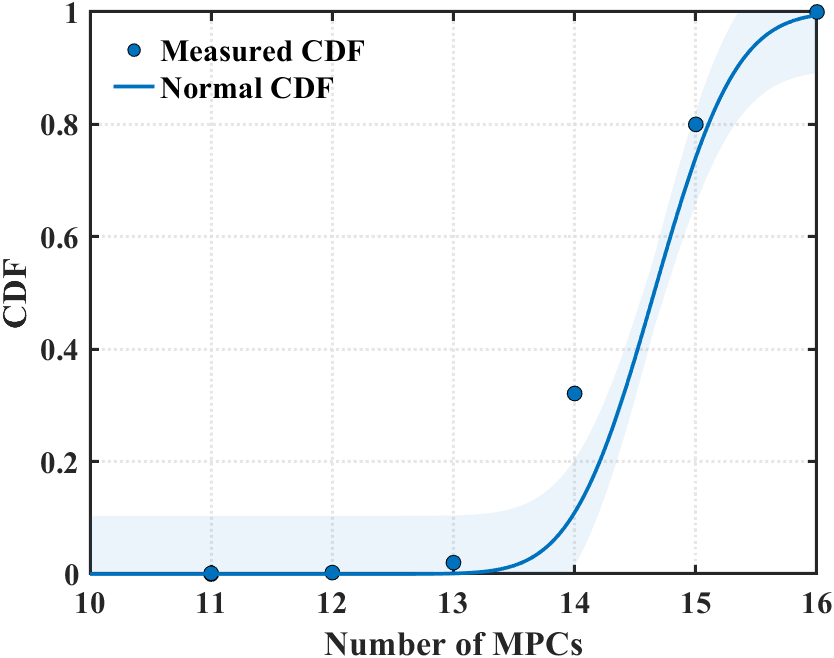}
	\end{minipage}
	
	\captionsetup{justification=raggedright,singlelinecheck=false}
	\caption{Statistical distributions of MPCs parameters.}
	\label{fig:canyon_compare}
\end{figure*}

For the small-scale, the analysis focuses on the key parameters required for statistical channel generation. According to the model framework established in Section~II, six parameters are extracted and analyzed, namely, the power, delay, angle of arrival (AoA), elevation of arrival (EOA), number of clusters, and number of multipath components within each cluster. These parameters jointly determine the statistical behavior of the generated channel in the power, delay, and spatial domains. 

For each snapshot, we run the SAGE algorithm~\cite{36,37} to estimate MPC parameters first, including delay, azimuth of arrival (AoA), and power. SAGE is a widely used high-resolution parameter estimation method under the expectation--maximization (EM) framework. The estimated MPCs of Intersection~2 are shown in Fig.~\ref{fig:MPC}. Each dot represents one MPC, and the color indicates the corresponding received power. Before the vehicle turns, most energy stays in a narrow AoA range aligned with the street. Around the turning point and the LOS--NLOS transition, the dominant AoA shifts and many components spread over a wider angular range. After entering the blocked street, the received power drops. The angular distribution becomes more dispersed and may contain multiple weak clusters. This indicates strong angular non-stationarity at intersections. Based on the extracted effective multipath components, the small-scale parameters are statistically analyzed under both LOS and NLOS link states. Since these parameters differ in physical meaning and random variation, suitable probability distributions are selected for each of them. The corresponding small-scale distributions and fitting results are shown in Fig.~\ref{fig:canyon_compare}, and their mathematical expressions are as follows:

\begin{itemize}
	
	\item \textbf{Normal distribution:}
	\begin{equation}
		f(x)=\frac{1}{\sigma\sqrt{2\pi}}
		\exp\left(-\frac{(x-\mu)^2}{2\sigma^2}\right)
	\end{equation}
	where $\mu$ is the mean and $\sigma^2$ is the variance.
	
	\item \textbf{Lognormal distribution:}
	\begin{equation}
		f(x)=\frac{1}{x\sigma\sqrt{2\pi}}
		\exp\left(-\frac{(\ln x-\mu)^2}{2\sigma^2}\right), \qquad x>0
	\end{equation}
	where $\mu$ is the mean and $\sigma^2$ is the variance of $\ln x$.
	
	\item \textbf{Laplace distribution:}
	\begin{equation}
		f(x)=\frac{1}{2b}
		\exp\left(-\frac{|x-\mu|}{b}\right)
	\end{equation}
	where $\mu$ is the location parameter and $b$ is the scale parameter.
	
\end{itemize}

The associated distribution parameters are further expressed as functions of the environmental factor~$S$, as listed in Tables~\ref{tab:los_statistical_parameters} and Tables~\ref{tab:nlos_statistical_parameters}. 
The fitting results of the multipath power parameter are shown in the Fig.~\ref{fig:semantic_fit}. For convenience of subsequent computation, the environmental factor is normalized as $\tilde S=(S-30)/15$.

For the LOS state, the fitted relationships between the small-scale parameters and the environmental factor are summarized in Table~II. It can be seen that different parameters follow different statistical distributions. The delay $\tau$ is better described by a lognormal distribution. The AoA $\theta$ and the EOA $\phi$ are both modeled by Laplace distributions. The relative power $\beta$, the number of clusters $N_{\mathrm{cl}}$, and the number of multipath components within each cluster $N_{\mathrm{MPC}}$ are fitted by normal distributions. In addition, both the location and dispersion parameters of these distributions vary with the environmental factor~$S$. Some of them can be well represented by linear functions, while others are better described by exponential forms for the dispersion term.

For the NLOS state, the fitted relationships between the small-scale parameters and the environmental factor are given in Table~III. Compared with the LOS case, the small-scale statistics under NLOS conditions show stronger sensitivity to the environment. Specifically, the delay $\tau$ is better modeled by a Laplace distribution in the NLOS state. The AoA $\theta$ and the EOA $\phi$ are still described by Laplace distributions. The relative power $\beta$, the number of clusters $N_{\mathrm{cl}}$, and the number of multipath components within each cluster $N_{\mathrm{MPC}}$ are again modeled by normal distributions. In comparison with the LOS case, multipath propagation is more complex under NLOS conditions. The combined effects of building blockage and corner propagation make the distribution range and dispersion of these parameters more sensitive to the variation of~$S$.

\section{Model Validation and Discussion}
In this section, the effectiveness of the proposed environment-related channel model is evaluated and future work is discussed. The accuracy of the large-scale path-loss model is verified by comparing the model predictions with the measured path-loss results. Then, for a given environmental factor $S$, propagation state $\xi$, and Tx--Rx distance $d$, the channel impulse response is generated based on the proposed model. Key statistical metrics are further extracted and compared with those obtained from measurements. Through these comparisons, the capability of the proposed model to characterize channel propagation in urban street-canyon intersections can be systematically assessed.

\begin{figure}[t]
	\centering
	\includegraphics[width=1\columnwidth]{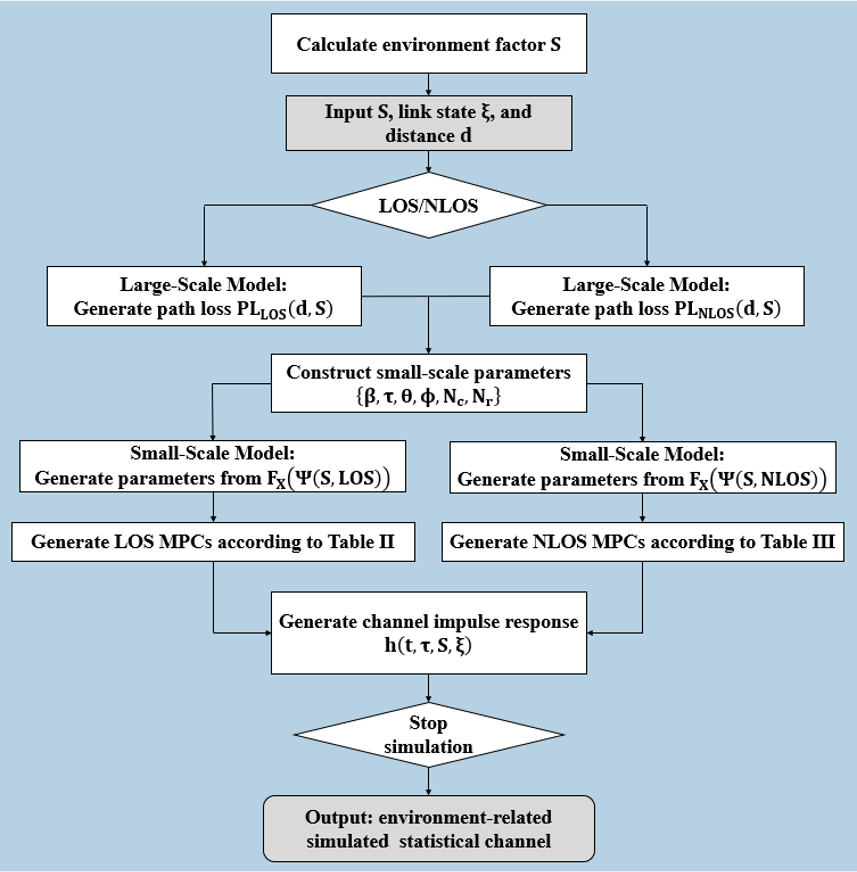}
	\caption{Flowchart of the implementation steps of the proposed environment-related intersection channel model.}
	\label{fig:procedure}
\end{figure}

\subsection{Model Implementation}
Based on the proposed environment-related model, the simulated channel is generated according to the procedure shown in Fig.~\ref{fig:procedure}. 

First, the composite environmental factor $S$ is calculated from the building-related information extracted from the intersection scenario. Then, $S$, the link state $\xi$, and the Tx--Rx distance $d$ are taken as inputs, and the current link is classified as either LOS or NLOS. At the large-scale level, the corresponding path-loss model is selected according to the link state, thereby determining the large-scale attenuation level for the current scenario. After the large-scale path loss is determined, the small-scale parameter set $\{\beta,\tau,\theta,\phi,N_c,N_r\}$ is constructed, where $\beta$, $\tau$, $\theta$, $\phi$, $N_c$, and $N_r$ denote the multipath power, propagation delay, angle of arrival, elevation of arrival, number of clusters, and number of multipath components within each cluster, respectively. According to the current link state, the small-scale parameters are generated from the corresponding conditional distribution, namely, $f_{\mathbf X}(\mathbf X\mid S,\mathrm{LOS})$ or $f_{\mathbf X}(\mathbf X\mid S,\mathrm{NLOS})$. Then, based on the parameter settings and statistical relationships listed in Table~II, the multipath components (MPCs) under LOS or NLOS conditions are generated, and the power, delay, and angular parameters of each cluster and its intra-cluster components are specified. Finally, all generated MPCs are superimposed to obtain the environment-dependent channel impulse response $h(t,\tau;S,\xi)$. Once the channel generation for the current scenario is completed, the simulation is terminated and the resulting environment-related statistical channel is obtained. Through this procedure, the proposed model enables statistical channel generation driven by the environmental factor $S$, such that the generated channel can jointly capture the large-scale attenuation characteristics and the small-scale multipath structure of urban street-canyon intersections.

\begin{figure*}[t]
	\centering
	% -------- Row 1: a1 b1 c1 --------
	\subfloat[]{%
		\includegraphics[width=0.33\linewidth]{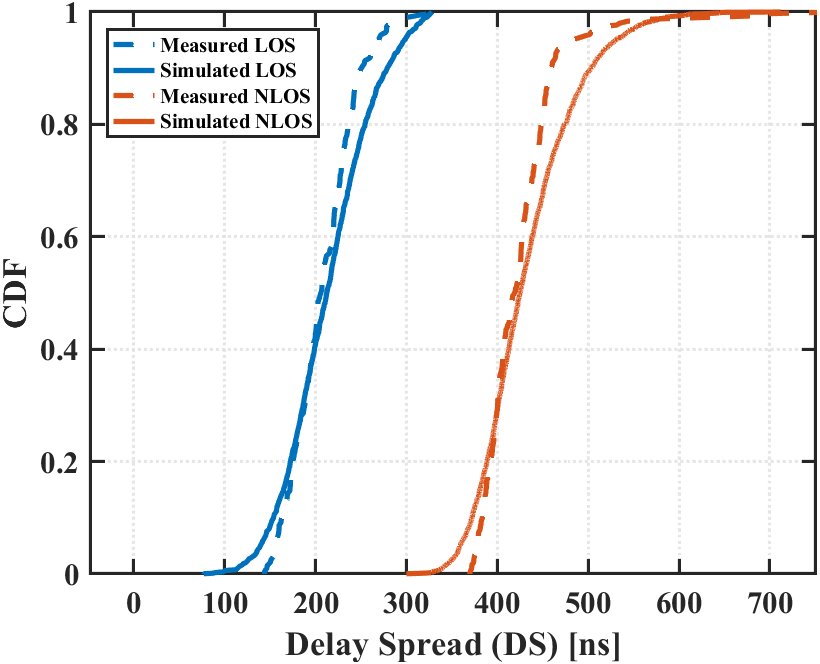}%
	}\hfill
	\subfloat[]{%
		\includegraphics[width=0.33\linewidth]{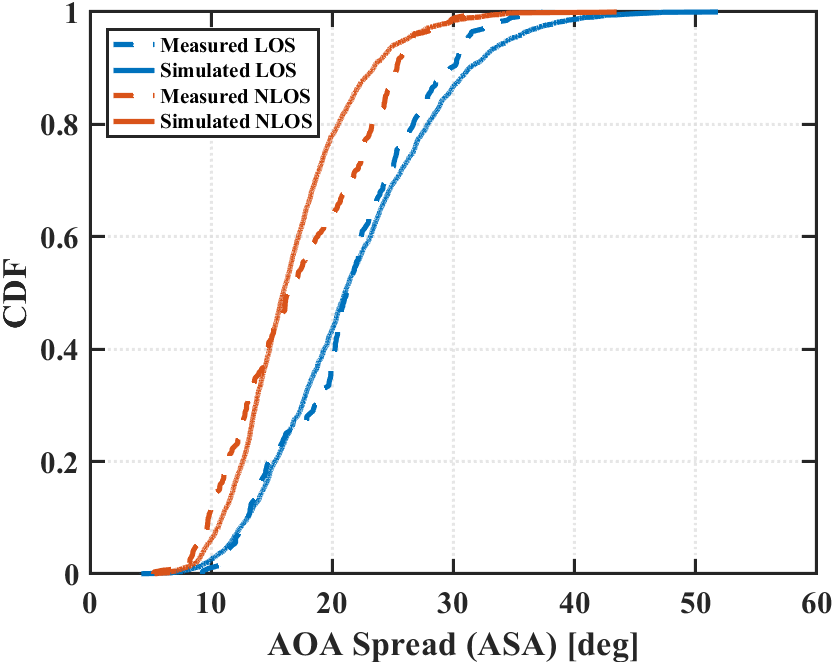}%
	}\hfill
	\subfloat[]{%
		\includegraphics[width=0.33\linewidth]{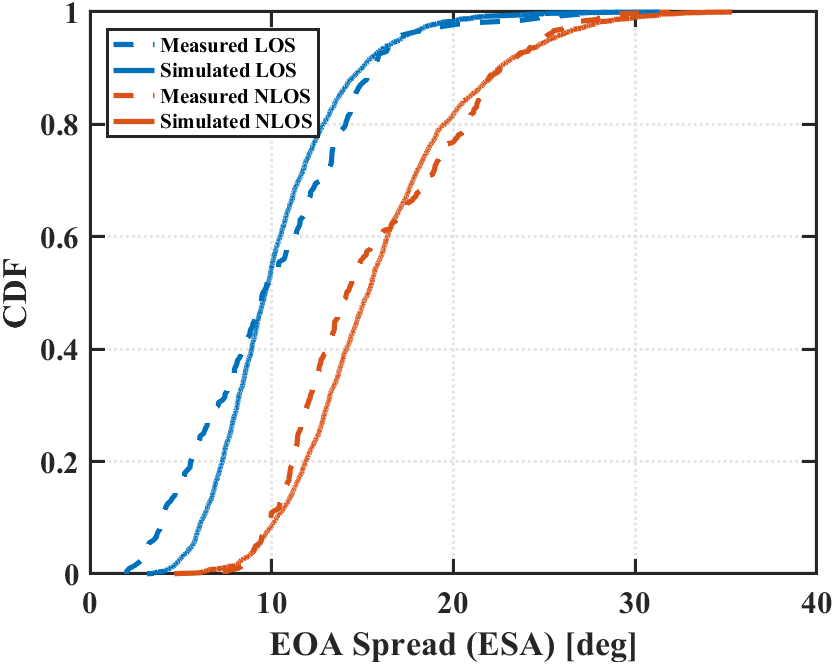}%
	}
    
    \captionsetup{justification=raggedright,singlelinecheck=false}
	\caption{Comparison between the proposed model and measured data. (a) DS; (b) ASA; (c) ESA.}
	
	\label{fig:six_plots}
\end{figure*}

\subsection{Model Validation and Future Work}
To validate the proposed model, a new measured medium-complexity intersection(MCL) is selected for testing. Two third-order statistical metrics are adopted, namely the root mean square (RMS) delay spread and the angular spreads of the angle of arrival (AoA) and elevation of arrival (EOA). These metrics are used to evaluate the proposed model. The RMS delay spread is calculated as the second central moment of the power delay profile (PDP)~\cite{Kanhere2025NYURay}, while the angular spreads are computed according to Fleury's~\cite{Fleury2000} definition. The Fig.~\ref{fig:six_plots} compares the measured data with the simulated results generated by the proposed model. It can be observed that the distributions of the generated delay spread and angular spread agree well with the measurements.

The above validation results indicate that the proposed environment-related statistical channel model achieves good accuracy and applicability in urban street-canyon intersection scenarios. Nevertheless, the model still has certain limitations. First, the current model is established based on 5.8~GHz measurements collected in urban street-canyon intersections. Its parameter forms and the weights of the environmental factor $S$ are mainly derived from the measured scenarios considered in this work. Further validation and extension are still needed for other frequency bands, other types of urban roads, and more complex dynamic scenarios. Second, the proposed model uses a single environmental factor $S$ to provide a compact representation of the local building environment. Although this factor can capture the major environmental differences, it still has limited capability in describing more detailed geometric features. Future work can extend the environmental descriptors and the parameter-mapping model based on richer measurement data, so as to improve the generalization ability of the model in a wider range of scenarios. In addition, extending the present framework toward semantic-aware channel characterization for emerging integrated sensing and communication scenarios is also a direction~\cite{Zhang2026semantic}.

\section{Conclusion}

This paper proposes an environment-related statistical channel model for urban street-canyon intersections based on 5.8~GHz channel measurements. To account for the differences in building height, variation, and density among different intersections, an environmental factor is constructed to provide a unified description of environmental impact on propagation. By combining the large-scale and small-scale models, environment-dependent statistical channels are generated, thereby establishing a link between physical intersection environment and statistical channel representation. Validation results show that the proposed method can effectively capture the main propagation characteristics of urban street-canyon intersections. The proposed model offers both physical interpretability and statistical modeling capability, and can serve as a useful tool for channel simulation, performance evaluation in urban vehicular wireless systems.

\vfill


\begin{thebibliography}{00}
	

	
	\bibitem{14}
	S.~Gyawali, S.~Xu, Y.~Qian, and R.~Q.~Hu,
	``Challenges and solutions for cellular based V2X communications,''
	\emph{IEEE Commun. Surveys Tuts.},
	vol.~23, no.~1, pp.~222--255, 1st Quart. 2021.
	
	\bibitem{18}
	R.~He \emph{et~al.},
	``Characterization of quasi-stationarity regions for vehicle-to-vehicle radio channels,''
	\emph{IEEE Trans. Antennas Propag.},
	vol.~63, no.~5, pp.~2237--2251, May 2015.
		
	\bibitem{He2020mmwave}
	R.~He \emph{et~al.},
	``Propagation channels of 5G millimeter-wave vehicle-to-vehicle communications: Recent advances and future challenges,''
	\emph{IEEE Veh. Technol. Mag.},
	vol.~15, no.~1, pp.~16--26, 2020.
	
	\bibitem{85}
	C.-X.~Wang, J.~Bian, J.~Sun, W.~Zhang, and M.~Zhang,
	``A survey of 5G channel measurements and models,''
	\emph{IEEE Commun. Surveys Tuts.},
	vol.~20, no.~4, pp.~3142--3168, 4th Quart. 2018.
	
	\bibitem{He2024book}
	R.~He and B.~Ai,
	\emph{Wireless Channel Measurement and Modeling in Mobile Communication Scenario: Theory and Application}.
	Boca Raton, FL, USA: CRC Press, 2024.
	
	\bibitem{86}
	A.~F.~Molisch, F.~Tufvesson, J.~Karedal, and C.~F.~Mecklenbr\"auker,
	``A survey on vehicle-to-vehicle propagation channels,''
	\emph{IEEE Wireless Commun.},
	vol.~16, no.~6, pp.~12--22, Dec. 2009.
	
	\bibitem{87}
	S.~Chen, J.~Hu, Y.~Shi, L.~Zhao, and W.~Li,
	``A vision of C-V2X: Technologies, field testing, and challenges with Chinese development,''
	\emph{IEEE Internet Things J.},
	vol.~7, no.~5, pp.~3872--3881, May 2020.
	
	\bibitem{He2013Cutting}
	R.~He, Z.~Zhong, B.~Ai, J.~Ding, Y.~Yang, and A.~F.~Molisch,
	``Short-term fading behavior in high-speed railway cutting scenario: Measurements, analysis, and statistical models,''
	\emph{IEEE Trans. Antennas Propag.},
	vol.~61, no.~4, pp.~2209--2222, Apr. 2013.
	
	\bibitem{51}
	Z.~Xu \emph{et~al.},
	``Relaying for IEEE 802.11p at road intersection using a vehicular non-stationary channel model,''
	in \emph{Proc. IEEE 6th Int. Symp. Wireless Veh. Commun. (WiVeC)},
	Vancouver, BC, Canada, 2014, pp.~1--6.
	
	\bibitem{Zheng2020nlos}
	Q.~Zheng, R.~He, B.~Ai, C.~Huang, W.~Chen, Z.~Zhong, and H.~Zhang,
	``Channel non-line-of-sight identification based on convolutional neural networks,''
	\emph{IEEE Wireless Commun. Lett.},
	vol.~9, no.~9, pp.~1500--1504, 2020.
	
	\bibitem{12}
	C.~Liu, R.~He, Y.~Niu, S.~Mao, B.~Ai, and R.~Chen,
	``Refracting reconfigurable intelligent surface assisted URLLC for millimeter wave high-speed train communication coverage enhancement,''
	\emph{IEEE Trans. Veh. Technol.},
	vol.~74, no.~1, pp.~953--967, Jan. 2025.
	
	\bibitem{Qian6G}
	Z.~Qian, Z.~Li, W.~Zhou, C.~Huang, and C.-X.~Wang,
	``6G wireless channel scenario extensions and characteristics analysis for urban environment,''
	in \emph{Proc. IEEE 97th Veh. Technol. Conf. (VTC2023-Spring)},
	Florence, Italy, 2023.
	
	\bibitem{TR38901}
	3GPP,
	\emph{Study on Channel Model for Frequencies From 0.5 to 100~GHz},
	3GPP TR~38.901, v16.1.0, Dec. 2019.
	
	\bibitem{WINNER2}
	P.~Ky\"{o}sti \emph{et~al.},
	\emph{WINNER II Channel Models},
	IST-WINNER II Deliverable D1.1.2, ver.~1.1, Sep. 2007.
	
	\bibitem{COST231}
	B.~S.~L.~Castro, I.~R.~Gomes, F.~C.~J.~Ribeiro, and G.~P.~S.~Cavalcante,
	``COST231-Hata and SUI models performance using a LMS tuning algorithm on 5.8~GHz in Amazon Region cities,''
	in \emph{Proc. 4th Eur. Conf. Antennas Propag. (EuCAP)},
	Barcelona, Spain, 2010.
	
	\bibitem{Zemen2025SSCR}
	T.~Zemen \emph{et~al.},
	``Site-specific radio channel representation for 5G and 6G,''
	\emph{IEEE Commun. Mag.},
	vol.~63, no.~6, pp.~106--113, 2025.
	
	\bibitem{Sai}
	S.~Sai, E.~Niwa, K.~Mase, M.~Nishibori, J.~Inoue, M.~Obuchi, T.~Harada,
	H.~Ito, K.~Mizutani, and M.~Kizu,
	``Field evaluation of UHF radio propagation for an ITS safety system in an urban environment,''
	\emph{IEEE Commun. Mag.},
	vol.~47, no.~11, pp.~120--127, 2009.
	
	\bibitem{Mur}
	J.~Muramatsu, N.~Suzuki, Y.~Ito, and T.~Taga,
	``Measurement of radio propagation characteristics for inter-vehicle communication in urban areas,''
	in \emph{Proc. Int. Symp. Antennas Propag. (ISAP)},
	Aug. 2007.
	
	\bibitem{Ito}
	Y.~Ito, T.~Taga, J.~Muramatsu, and N.~Suzuki,
	``Prediction of line-of-sight propagation loss in inter-vehicle communication environments,''
	in \emph{Proc. IEEE 18th Int. Symp. Pers., Indoor, Mobile Radio Commun. (PIMRC)},
	2007, pp.~1--5.
	
	\bibitem{Gus}
	C.~Gustafson, K.~Mahler, D.~Bolin, and F.~Tufvesson,
	``The COST IRACON geometry-based stochastic channel model for vehicle-to-vehicle communication in intersections,''
	\emph{IEEE Trans. Veh. Technol.},
	vol.~69, no.~3, pp.~2365--2375, 2020.
	
	\bibitem{Liu}
	Y.~Liu \emph{et~al.},
	``Channel measurements and characteristics analysis for suburban scenarios at 10~GHz,''
	in \emph{Proc. IEEE/CIC Int. Conf. Commun. China (ICCC)},
	Shanghai, China, 2025, pp.~1--5.
	
	\bibitem{Rad}
	J.~Karedal, F.~Tufvesson, T.~Abbas, O.~Klemp, A.~Paier, L.~Bernad\'o, and A.~F.~Molisch,
	``Radio channel measurements at street intersections for vehicle-to-vehicle safety applications,''
	in \emph{Proc. IEEE 71st Veh. Technol. Conf. (VTC)},
	2010, pp.~1--5.
	
	\bibitem{Eva}
	K.~Mahler, P.~Paschalidis, M.~Wisotzki, A.~Kortke, and W.~Keusgen,
	``Evaluation of vehicular communication performance at street intersections,''
	in \emph{Proc. IEEE 80th Veh. Technol. Conf. (VTC Fall)},
	2014, pp.~1--5.
	
	\bibitem{pat}
	M.~Nilsson, C.~Gustafson, T.~Abbas, and F.~Tufvesson,
	``A path loss and shadowing model for multilink vehicle-to-vehicle channels in urban intersections,''
	\emph{Sensors},
	vol.~18, no.~12, 2018.
	
	\bibitem{Del}
	L.~Bernad\'o, T.~Zemen, F.~Tufvesson, A.~F.~Molisch, and C.~F.~Mecklenbr\"auker,
	``Delay and Doppler spreads of nonstationary vehicular channels for safety-relevant scenarios,''
	\emph{IEEE Trans. Veh. Technol.},
	vol.~63, no.~1, pp.~82--93, 2014.
	
	\bibitem{17}
	B.~Ai \emph{et~al.},
	``Challenges toward wireless communications for high-speed railway,''
	\emph{IEEE Trans. Intell. Transp. Syst.},
	vol.~15, no.~5, pp.~2143--2158, Oct. 2014.
	
	\bibitem{16}
	M.~Yang \emph{et~al.},
	``Measurements and cluster-based modeling of vehicle-to-vehicle channels with large vehicle obstructions,''
	\emph{IEEE Trans. Wireless Commun.},
	vol.~19, no.~9, pp.~5860--5874, Sep. 2020.
	
	\bibitem{19}
	R.~He \emph{et~al.},
	``Clustering enabled wireless channel modeling using big data algorithms,''
	\emph{IEEE Commun. Mag.},
	vol.~56, no.~5, pp.~177--183, May 2018.
	
	\bibitem{Bai2025RIS}
	D.~Bai, S.~Yang, Y.~Wang, J.~Huang, C.-X.~Wang, and F.-C.~Zheng,
	``Ray tracing channel modeling for 6G RIS-beamforming communications at 28 GHz,''
	in \emph{Proc. IEEE 102nd Veh. Technol. Conf. (VTC2025-Fall)},
	Chengdu, China, 2025, pp.~1--5.
	
	\bibitem{5}
	Z.~Su \emph{et~al.},
	``Analysis of channel non-stationarity for V2V and V2I communications at 5.9~GHz in urban scenarios,''
	in \emph{Proc. IEEE Int. Conf. Commun. Workshops (ICC Workshops)},
	Seoul, South Korea, 2022, pp.~1130--1134.
	
	\bibitem{55}
	L.~Tian, J.~Zhang, H.~Tan, P.~Tang, and G.~Liu,
	``Propagation characteristics of elevation angles and three-dimensional fading channel model with angle offset,''
	\emph{China Commun.},
	vol.~16, no.~9, pp.~62--78, Sep. 2019.
	
	\bibitem{35}
	D.~Zhao, C.~Huang, C.-X.~Wang, and J.~Li,
	``Scenario classification and channel modeling for MIMO communications in suburban road scenarios,''
	in \emph{Proc. 18th Eur. Conf. Antennas Propag. (EuCAP)},
	Glasgow, U.K., 2024, pp.~1--5.
	
	\bibitem{58}
	Z.~Zhang \emph{et~al.},
	``A general channel model for integrated sensing and communication scenarios,''
	\emph{IEEE Commun. Mag.},
	vol.~61, no.~5, pp.~68--74, May 2023.
	
	\bibitem{9}
	G.~Zhang, X.~Cai, N.~Franchi, and M.~L{\"u}bke,
	``A geometry-based ISAC channel model for vehicle-to-vehicle scenarios,''
	in \emph{Proc. 19th Eur. Conf. Antennas Propag. (EuCAP)},
	Stockholm, Sweden, 2025, pp.~1--5.
	
	\bibitem{61}
	C.~Huang, R.~Wang, P.~Tang, R.~He, B.~Ai, Z.~Zhong, C.~Oestges, and A.~F.~Molisch,
	``Geometry-cluster-based stochastic MIMO model for vehicle-to-vehicle communications in street canyon scenarios,''
	\emph{IEEE Trans. Wireless Commun.},
	vol.~20, no.~2, pp.~755--770, 2020.
	
	\bibitem{Zhang2023DT}
	Y.~Zhang \emph{et~al.},
	``Generative adversarial networks based digital twin channel modeling for intelligent communication networks,''
	\emph{China Commun.},
	vol.~20, no.~8, pp.~32--43, 2023.
	
	\bibitem{He2026AI}
	R.~He, M.~Yang, Z.~Zhang, B.~Ai, and Z.~Zhong,
	``Artificial intelligence empowered channel prediction: A new paradigm for propagation channel modeling,''
	\emph{arXiv preprint arXiv:2601.09205}, 2026.
	
	\bibitem{He2025INTERACT}
	R.~He, N.~D.~Cicco, B.~Ai, M.~Yang, Y.~Miao, and M.~Boban,
	``COST CA20120 INTERACT framework of artificial intelligence-based channel modeling,''
	\emph{IEEE Wireless Commun.},
	vol.~32, no.~4, pp.~200--207, 2025.
	
	\bibitem{yang}
	M.~Yang \emph{et~al.},
	``Measurement and characterization of vehicle-to-vehicle channels in vegetated environment,''
	\emph{IEEE Trans. Veh. Technol.},
	vol.~73, no.~11, pp.~15955--15968, Nov. 2024.
	
	\bibitem{36}
	J.~A.~Fessler and A.~O.~Hero,
	``Space-alternating generalized expectation-maximization algorithm,''
	\emph{IEEE Trans. Signal Process.},
	vol.~42, no.~10, pp.~2664--2677, Oct. 1994.
	
	\bibitem{37}
	M.~Matthaiou, D.~I.~Laurenson, N.~Razavi-Ghods, and S.~Salous,
	``Characterization of an indoor MIMO channel in frequency domain using the 3D-SAGE algorithm,''
	in \emph{Proc. IEEE Int. Conf. Commun. (ICC)},
	Glasgow, U.K., 2007, pp.~5868--5872.
	
	\bibitem{Kanhere2025NYURay}
	O.~Kanhere, H.~Poddar, and T.~S.~Rappaport,
	``Calibration of NYURay for ray tracing using 28, 73, and 142~GHz channel measurements conducted in indoor, outdoor, and factory scenarios,''
	\emph{IEEE Trans. Antennas Propag.},
	vol.~73, no.~1, pp.~405--420, Jan. 2025.
	
	\bibitem{Fleury2000}
	B.~H.~Fleury,
	``First- and second-order characterization of direction dispersion and space selectivity in the radio channel,''
	\emph{IEEE Trans. Inf. Theory},
	vol.~46, no.~6, pp.~2027--2044, Sep. 2000.
	
	\bibitem{Zhang2026semantic}
	Z.~Zhang, R.~He, B.~Ai, M.~Yang, X.~Zhang, Z.~Qi, and Z.~Zhong,
	``Channel semantic characterization for integrated sensing and communication scenarios: From measurements to modeling,''
	\emph{IEEE Trans. Wireless Commun.},
	vol.~25, pp.~13073--13089, 2026.
	
	
	
	
	
     
\end{thebibliography}
\end{document}